\documentclass[12pt,leqno]{amsart}
\usepackage{amssymb}
\usepackage{amsmath,amssymb}
\newtheorem{proposition}{Proposition}[section]

\begin{document}
\theoremstyle{plain}
\newtheorem{MainThm}{Theorem}
\newtheorem{thm}{Theorem}[section]
\newtheorem{clry}[thm]{Corollary}
\newtheorem{prop}[thm]{Proposition}
\newtheorem{lem}[thm]{Lemma}
\newtheorem{deft}[thm]{Definition}
\newtheorem{hyp}{Assumption}
\newtheorem*{ThmLeU}{Theorem (J.~Lee, G.~Uhlmann)}

\theoremstyle{definition}
\newtheorem{rem}[thm]{Remark}
\newtheorem*{acknow}{Acknowledgments}
\numberwithin{equation}{section}
\newcommand{\eps}{{\varphi}repsilon}
\renewcommand{\d}{\partial}
\newcommand{\re}{\mathop{\rm Re} }
\newcommand{\im}{\mathop{\rm Im}}
\newcommand{\R}{\mathbf{R}}
\newcommand{\C}{\mathbf{C}}
\newcommand{\N}{\mathbf{N}}
\newcommand{\D}{C^{\infty}_0}
\renewcommand{\O}{\mathcal{O}}
\newcommand{\dbar}{\overline{\d}}
\newcommand{\supp}{\mathop{\rm supp}}
\newcommand{\abs}[1]{\lvert #1 \rvert}
\newcommand{\csubset}{\Subset}
\newcommand{\detg}{\lvert g \rvert}
\newcommand{\ppp}{\partial}
\newcommand{\dd}{\mbox{div}\thinspace}
\newcommand{\www}{\widetilde}

\title
[Calder\'on problem for Maxwell's equations]
{Calder\'on problem for Maxwell's equations in the wave guide}

\author{
O.~Yu.~Imanuvilov and \,
M.~Yamamoto }
\thanks{ Department of Mathematics, Colorado State
University, 101 Weber Building, Fort Collins, CO 80523-1874, U.S.A.
e-mail: {\tt oleg@math.colostate.edu}.
Partially supported by NSF grant DMS 1312900}\,
\thanks{ Department of Mathematical Sciences, The University
of Tokyo, Komaba, Meguro, Tokyo 153, Japan e-mail:
myama@ms.u-tokyo.ac.jp}

\date{}

\maketitle

\begin{abstract}
For Maxwell's equations in a wave guide, we prove the global uniqueness
in determination of
the conductivity, the permeability and the permittivity
by partial Dirichlet-to-Neumann
map limited to an arbitrary subboundary.
\end{abstract}

Let $\www\Omega$ be a cylinder in $ \R^3$  and let $i = \sqrt{-1}$, $x = (x_1,x_2,x_3) \in \R^3$.
Let $\www E=(\www E_1,\www E_2,\www E_3)$ be the electric field,
$\www H=(\www H_1,\www H_2,\www H_3)$
the magnetic field, $\sigma$ be the conductivity, $\mu$ the permeability
and $\epsilon$ permittivity and $\omega\in \R^1, \omega\ne 0$ be a frequency.
Then Maxwell's equations are given by
\begin{equation}\label{klinok1}
\mbox{curl}\,\www E-i\omega\mu \www H=0,\quad\mbox{in}\,\,\www\Omega,
\end{equation}
\begin{equation}\label{klinok2}
\mbox{curl}\,\www H+i\omega\gamma \www E =0,\quad\mbox{in}\,\,\www\Omega.
\end{equation}
In this paper, we consider Maxwell's equation inside a wave guide $\www\Omega$
(e.g., Jackson \cite{J}, Chapter 8).
More precisely, let $\Omega \subset \R^2$ be a bounded domain with smooth
boundary $\partial\Omega$ and $\www\Omega = \Omega \times (-\infty,\infty)
= \{ (x_1,x_2,x_3); \thinspace (x_1,x_2) \in \Omega, \thinspace
x_3 \in \R\}.$  We assume
$$
\www E(x_1,x_2,x_3) = E(x_1,x_2)e^{hx_3}, \quad
\www H(x_1,x_2,x_3) = H(x_1,x_2)e^{hx_3}, \quad x_3\in \R^1,
$$
where $h\in \C$ is a constant.
Then we can rewrite Maxwell's equations in $\Omega$:
\begin{equation}\label{3}
L_{1,\mu,\gamma}(x,D)(E,H):=\left ( \begin{matrix}
\partial_{x_2}E_3-hE_2\\
-\partial_{x_1}E_3+hE_1\\\partial_{x_1}E_2-\partial_{x_2}E_1   \end{matrix}
\right)-i\omega\mu\left (\begin{matrix}H_1\\H_2\\
H_3\end{matrix} \right)=0,\quad\mbox{in}\,\,\Omega,
\end{equation}
and
\begin{equation}\label{4}
L_{2,\mu,\gamma}(x,D)(E,H):=\left ( \begin{matrix} \partial_{x_2}H_3-hH_2\\
-\partial_{x_1}H_3+hH_1\\\partial_{x_1}H_2-\partial_{x_2}H_1   \end{matrix}
\right)+i\omega\gamma\left (\begin{matrix}E_1\\E_2\\
E_3\end{matrix} \right)=0,\quad\mbox{in}\,\,\Omega.
\end{equation}
Here and henceforth we set
$$
\gamma=\epsilon+\frac{i\sigma}{\omega}
$$
and
\begin{equation}\label{zoika}
L_{\mu,\gamma}(x,D)(E,H) = (L_{1,\mu,\gamma}(x,D)(E,H),
L_{2,\mu,\gamma}(x,D)(E,H)).
\end{equation}
By $(\nu_1,\nu_2)$ we denote the outward unit normal vector to
$\partial\Omega$ and we set $\vec{\nu} = (\nu_1, \nu_2, 0)$.
Then we note that
\begin{equation}\label{nona1}
{\vec \nu\times E}
= \left ( \begin{matrix} \nu_2E_3\\
-\nu_1E_3\\
\nu_1E_2 - \nu_2E_1
\end{matrix}\right)  \quad \mbox{on $\partial\Omega .$}
\end{equation}

Let $\widetilde \Gamma$ be some fixed open subset of $\partial\Omega$ and
$\Gamma_0=\partial\Omega\setminus\widetilde \Gamma.$
Consider the following Dirichlet-to-Neumann map
\begin{equation}\label{2}
\Lambda_{\mu,\gamma}f = {\vec\nu}\times H
= \left (\begin{matrix} \nu_2H_3\\-\nu_1H_3\\
\nu_1H_2-\nu_2H_1
\end{matrix}\right)\quad\mbox{on}\,\,\widetilde \Gamma,
\end{equation}
where
$$
L_{\mu,\gamma}(x,D) (E,H)=0 \quad\mbox{in}\,\,\Omega,\quad
{\vec \nu}\times E\vert_{\Gamma_0}=0, \quad {\vec \nu}
\times E\vert_{\widetilde \Gamma}=f.
$$

In general for some  values of the parameter $\omega$, the boundary value problem
\begin{equation}\label{general}
L_{\mu,\gamma}(x,D) (E,H)=0\quad\mbox{in}\,\,\Omega,\quad \vec
\nu\times E\vert_{\Gamma_0}=0, \quad \vec \nu\times E\vert
_{\widetilde{\Gamma}}=f
\end{equation}
may not have a solution for some $f$.
By $D_{\mu,\gamma}$ we denote the set of functions
$f\in W^1_2(\widetilde\Gamma)$ such that there exists at least one solution
to (\ref{general}).  As for the mathematical theory on the boundary
value problem for Maxwell's equations, we refer for example to Dautray and
Lions \cite{DL}.

In general for some $f\in D_{\mu,\gamma}$, there exists more than one
solutions.  In that case as the value of $\Lambda_{\mu,\gamma}f$,
we consider the set of all  functions $\vec\nu\times H$ where the
pairs $(E,H)$ are the all possible solutions to (\ref{general}).
Thus our definition of the Dirichlet-to-Neumann map is different form
the classical one, and we have to specify the conception of the equality of
the Dirichlet-to-Neumann maps.

{\bf Definition.}
{\it We say that the Dirichelt-to-Neumann maps $\Lambda_{\mu_1,\gamma_1}$
and $\Lambda_{\mu_2,\gamma_2}$ are equal
if $D_{\mu_1,\gamma_1}\subset D_{\mu_2,\gamma_2}$
and for any pair $(E,H)$ which solves
\begin{equation}\nonumber
L_{\mu_1,\gamma_1}(x,D) (E,H)=0\quad\mbox{in}\,\,\Omega,\quad \vec \nu
\times E\vert_{\Gamma_0}=0, \quad \vec \nu\times E\vert_{\widetilde\Gamma}
= f,
\end{equation}
there exists a pair $(\widetilde E,\widetilde H)$ which solves
\begin{equation}\nonumber
L_{\mu_2,\gamma_2}(x,D) (\widetilde E,\widetilde H)=0
\quad\mbox{in}\,\,\Omega,\quad \vec \nu\times \tilde E\vert_{\Gamma_0}=0,
\quad \vec \nu\times\tilde  E\vert_{\widetilde\Gamma}=f
\end{equation}
and
$$
\vec\nu\times H=\vec\nu\times \tilde H\quad \mbox{on}\quad
{\widetilde\Gamma}.
$$
}

Then we can state our main result:
\\
\vspace{0.2cm}
\\
{\bf Theorem}
{\it
We assume that $h^2 + \omega^2\gamma_j\mu_j \ne 0$ on
$\overline{\Omega}$, $j=1,2$.
Let $\Omega$ be  a simply connected domain,  $\mu_j,\epsilon_j,\sigma_j\in C^5(\bar \Omega)$ for $j\in\{1,2\}$
and $\mu_j, \epsilon_j$ be the positive functions on $\bar \Omega.$
Suppose that $\Lambda_{\mu_1,\gamma_1} = \Lambda_{\mu_2,\gamma_2}$ and
$$
\mu_1-\mu_2=\gamma_1-\gamma_2=0\quad \mbox{on}\,\,\tilde \Gamma.
$$
Then $\mu_1=\mu_2$, $\epsilon_1 = \epsilon_2$ and $\sigma_1=\sigma_2$
in $\Omega$.
}

See Caro, Ola and Salo \cite{POM} and Ola, P\" aiv\"arinta and Somersalo
\cite{OPS} for the uniqueness results for Maxwell's equations in three
dimensions.  In a special two dimensional case of $h=0$, we refer to
Imanuvilov and Yamamoto \cite{IY2}, where
we can reduce Maxwell's equations
to the conductivity equation with zeroth order term and apply the
uniqueness result in Imanuvilov, Uhlmann and Yamamoto \cite{IUY} to
prove the theorem.

{\bf Notations.}
Let $i=\sqrt{-1}$ and $\overline{z}$ be the complex conjugate of
$z \in \Bbb C$. We set
$\partial_z = \frac 12(\partial_{x_1}-i\partial_{x_2})$,
$\partial_{\overline z}= \frac12(\partial_{x_1}+i\partial_{x_2})$.
For any holomorphic function $\Phi$ we set $\Phi'=\partial_z\Phi$ and $\bar\Phi'=\partial_{\bar z}\bar \Phi,$ $\Phi''= \partial^2_z\Phi$,$\bar\Phi''=\partial^2_{\bar z}\bar \Phi,$ $\vec e_1=(1,0),\vec e_2=(0,1).$  Let $\vec \tau=(\nu_2,-\nu_1)$ be tangential vector to $\partial\Omega$. Let $W^{1,\tau}_2(\Omega)$ be  the Sobolev space $W^1_2(\Omega)$ with the norm $\Vert u\Vert_{W^{1,\tau}_2(\Omega)}=\Vert \nabla u\Vert_{L^2(\Omega)}+\vert\tau\vert\Vert u\Vert_{L^2(\Omega)}.$ Moreover by $\lim_{\eta\to\infty} \frac{\Vert f(\eta)\Vert_X}{\eta} = 0$
and $\Vert f(\eta)\Vert_X \le C\eta$ as $\eta \to \infty$ with some $C>0$,
we define $f(\eta) = o_X(\eta)$ and $f(\eta) = O_X(\eta)$ as
$\eta \to \infty$ for a normed space $X$ with norm $\Vert \cdot\Vert_X$,
respectively.
\section{\bf Step 1:
Reduction of the Maxwell's equations \newline to the decoupled system of elliptic equations.}

 In this section we transform the system (\ref{3}), (\ref{4}) to some second order system of elliptic equation decoupled respect to the principal part.  Then we set up some new Dirichlet-to-Neumann  associated  with this system of elliptic equations and show that if $\Lambda_{\mu,\gamma}$ is known this Dirichlet-to-Neumann map is also known. From (\ref{3}) and (\ref{4}) we obtain
\begin{equation}\label{8}
H_1=\frac{1}{i\omega\mu}(\partial_{x_2}E_3-hE_2),\quad H_2=-\frac{1}{i\omega\mu}(\partial_{x_1}E_3-hE_1),\quad \partial_{x_1}H_2-\partial_{x_2}H_1=-i\omega\gamma E_3\quad\mbox{in}\,\,\Omega.
\end{equation}
and

\begin{equation}\label{9}
E_1=-\frac{1}{i\omega\gamma}(\partial_{x_2}H_3-hH_2),\quad E_2=\frac{1}{i\omega\gamma}(\partial_{x_1}H_3-hH_1),\quad \partial_{x_1}E_2-\partial_{x_2}E_1=i\omega\mu H_3\quad\mbox{in}\,\,\Omega.
\end{equation}
Denote $E'=(E_1,E_2), H'=(H_1,H_2).$
After we plug  into the third equation of (\ref{8}) expressions for $H_1$ and $H_2$ from the first two equations we obtain
\begin{equation}\label{10}
\mbox{div}\,\left (\frac{1}{i\omega\mu}\nabla E_3\right  )-h\mbox{div}\, (\frac {E'}{i\omega\mu})-i\omega\gamma E_3=0\quad\mbox{in}\,\,\Omega.
\end{equation}

Similarly, from (\ref{9}) we obtain

\begin{equation}\label{11}
\mbox{div}\,\left (\frac{1}{i\omega \gamma}\nabla H_3\right)-h\mbox{div}\, (\frac {H'}{i\omega\gamma})-i\omega\mu H_3=0\quad\mbox{in}\,\,\Omega.
\end{equation}
 We rewrite equations (\ref{8}) and (\ref{9}) as
\begin{equation}\label{granit1}
\left\{\begin{matrix}H_1+\frac{hE_2}{i\omega\mu}=\frac{1}{i\omega\mu}\partial_{x_2}E_3,\quad H_2-\frac{hE_1}{i\omega\mu}=-\frac{1}{i\omega\mu}\partial_{x_1}E_3\\
E_1-\frac{hH_2}{i\omega\gamma}=-\frac{1}{i\omega\gamma}\partial_{x_2}H_3,\quad E_2+\frac{hH_1}{i\omega\gamma}=\frac{1}{i\omega\gamma}\partial_{x_1}H_3
\end{matrix}\right.
\end{equation}
Solving the linear system (\ref{granit1}) respect to $E_1,E_2,H_1,H_2$ we have
\begin{equation}\label{granit2}\left\{\begin{matrix}
(1+\frac{h^2}{\omega^2\gamma\mu})E_2=\frac{h}{\omega^2\gamma\mu}\partial_{x_2}E_3+\frac{1}{i\omega\gamma}\partial_{x_1}H_3,
\\
(1+\frac{h^2}{\omega^2\gamma\mu})H_1=\frac{h}{\omega^2\gamma\mu}\partial_{x_1}H_3+\frac{1}{i\omega\mu}\partial_{x_2}E_3,
\\
(1+\frac{h^2}{\omega^2\gamma\mu})E_1=\frac{h}{\omega^2\gamma\mu}\partial_{x_1}E_3-\frac{1}{i\omega\gamma}\partial_{x_2}H_3,
\\
(1+\frac{h^2}{\omega^2\gamma\mu})H_2=\frac{h}{\omega^2\gamma\mu}\partial_{x_2}H_3-\frac{1}{i\omega\mu}\partial_{x_1}E_3.
\end{matrix}\right.
\end{equation}

We set $g=(1+\frac{h^2}{\omega^2\gamma\mu})$ and   $\nabla ^\bot P=(\frac{\partial P}{\partial x_2},-\frac{\partial P}{\partial x_1}).$

We can rewrite these equations as
\begin{equation}\label{golubka1}
E'=\frac{h}{\omega^2g\gamma\mu}\nabla E_3-\frac{1}{i\omega\gamma g}\nabla^\bot H_3,
\end{equation}

\begin{equation}\label{golubka2}
H'=\frac{h}{\omega^2g\gamma\mu}\nabla H_3+\frac{1}{i\omega\mu g}\nabla^\bot E_3,
\end{equation}

The simple computations imply

\begin{eqnarray}\label{oh1}h\mbox{div}\, (\frac {E'}{i\omega\mu})=\mbox{div}\,(\frac{h^2}{i\omega^3g\gamma\mu^2}\nabla E_3+\frac{h}{\omega^2\gamma\mu g}\nabla^\bot H_3)=
\nonumber\\
(\nabla\frac{h^2}{i\omega^3g\gamma\mu^2},\nabla E_3)+(\nabla\frac{h}{\omega^2\gamma\mu g},\nabla^\bot H_3)+\frac{h^2}{i\omega^3g\gamma\mu^2}\Delta E_3
\end{eqnarray}

and
\begin{eqnarray}\label{oh2}
h\mbox{div}\, (\frac {H'}{i\omega\gamma})=\mbox{div}(\frac{h^2}{i\omega^3g\gamma^2\mu}\nabla H_3-\frac{h}{\omega^2\mu\gamma g}\nabla^\bot E_3)=
\nonumber\\
(\nabla\frac{h^2}{i\omega^3g\gamma^2\mu},\nabla H_3)-(\nabla\frac{h}{\omega^2\mu\gamma g},\nabla^\bot E_3)+\frac{h^2}{i\omega^3g\gamma^2\mu}\Delta H_3.
\end{eqnarray}
We substitute the formulae for $h\mbox{div}\, (\frac {E'}{i\omega\mu})$ in (\ref{10}) and formulae for $h\mbox{div}\, (\frac {H'}{i\omega\gamma})$ into (\ref{11}) to obtain:

\begin{equation}\label{vizir}L_{(1)}(x,D)E_3=\mbox{div}\,\left (\left(\frac{1}{i\omega\mu}-\frac{h^2}{i\omega^3g\gamma\mu^2}\right)\nabla E_3\right  )-(\nabla\frac{h}{\omega^2\gamma\mu g},\nabla^\bot H_3)-i\omega\gamma E_3=0\quad\mbox{in}\,\,\Omega,
\end{equation}
\begin{equation}\label{lobster1}
E_3\vert_{\Gamma_0}=0
\end{equation}

and
\begin{equation}\label{11!}
L_{(2)}(x,D)H_3=\mbox{div}\,\left (\left (\frac{1}{i\omega \gamma}-\frac{h^2}{i\omega^3g\gamma^2\mu}\right )\nabla H_3\right)+(\nabla\frac{h}{\omega^2\mu\gamma g},\nabla^\bot E_3)-i\omega\mu H_3=0\quad\mbox{in}\,\,\Omega,
\end{equation}

\begin{equation}\label{p32}
\frac{\partial H_3}{\partial\nu}\vert_{\Gamma_0}=0.
\end{equation}

We set $\rho=(\rho_1,\rho_2,\rho_3)$ where
\begin{equation}\label{gnom}
\rho_1 = \frac{1}{i\omega\mu}-\frac{h^2}{i\omega^3g\gamma\mu^2}
= \frac{\omega\gamma}{i(h^2+\omega^2\gamma\mu)},
\end{equation}
$$
\rho_2=\frac{1}{i\omega \gamma}-\frac{h^2}{i\omega^3g\gamma^2\mu}
= \frac{\omega\mu}{i(h^2+\omega^2\gamma\mu)},
\quad \rho_3=\frac{h}{\omega^2\mu\gamma g}.
$$
Then we write (\ref{vizir}) -(\ref{p32}) as
\begin{equation}\label{granit3}
L(x,D)W=\Delta W+2A\partial_zW+2B\partial_{\bar z}W+QW=0\quad \mbox{in}\,\,\Omega,\quad \mathcal B(x,D) W\vert_{\Gamma_0}=0,
\end{equation}

where
\begin{equation}\label{megaphon}
A=\left ( \begin{matrix}\partial_{\bar z}\ln \rho_1& -\frac{i}{\rho_1}\partial_{\bar z}\rho_3\\
\frac{i}{\rho_2}\partial_{\bar z}\rho_3&\partial_{\bar z}\ln \rho_2
\end{matrix}\right),\quad B=\left ( \begin{matrix}\partial_{z}\ln \rho_1&\frac{ i}{\rho_1}\partial_{ z}\rho_3\\
-\frac{ i}{\rho_2}\partial_{ z}\rho_3&\partial_{ z}\ln \rho_2
\end{matrix}\right),\quad Q=\left ( \begin{matrix}\frac{-i\omega\gamma}{\rho_1}&0\\ 0&\frac{-i\omega\mu}{\rho_2}\end{matrix}\right),
\end{equation}
$
W=(E_3,H_3)$ and $\mathcal B(x,D) W=(E_3,\frac{\partial H_3}{\partial\nu}).$

Indeed the equation $E_3\vert_{\Gamma_0}=0$ follows from (\ref{nona1}) and (\ref{general}).  Let us show that the boundary condition (\ref{p32}) holds true.  By (\ref{nona1}),  (\ref{general}), (\ref{9})   and the boundary condition $\vec \nu\times E\vert_{\Gamma_0}=0$ we have on $\Gamma_0:$
\begin{equation}\label{lobster}
0=\nu_1E_2 - \nu_2E_1=\nu_1\frac{1}{i\omega\gamma}(\partial_{x_1}H_3-hH_1)+\nu_2\frac{1}{i\omega\gamma}(\partial_{x_2}H_3-hH_2)=\frac{1}{i\omega\gamma}(\partial_\nu H_3-h(\vec \nu,H')).
\end{equation}

From (\ref{8}) we have
$$(\vec\nu,H')=\frac{1}{i\omega\mu}\partial_{\vec\tau}E_3-h\nu_1E_2+h\nu_2 E_1=0\quad \mbox{on}\,\,\Gamma_0.
$$
Hence (\ref{lobster1}) and  (\ref{nona1}), (\ref{general}) imply (\ref{p32}).

We set $\mathcal R(x,D)W=(\partial_\nu W_1,W_2).$

\begin{proposition}\label{leika}
Let the traces of the functions $\gamma$ and $\mu$ are given on
$\tilde \Gamma.$ If the Dirichlet-to-Neumann map (\ref{2}) is known,
then the  following Dirichelt-to-Neumann map is known:
\begin{equation}\label{kinovolt}
\mbox{\bf f}\rightarrow \mathcal R(x,D)W\vert_{\tilde\Gamma},\quad L(x,D)W=0\quad\mbox{in}\,\,\Omega, \quad \mathcal B(x,D)W=\mbox{\bf f},\quad\mbox{supp}\, \mbox{\bf f}\subset\tilde\Gamma.
\end{equation}
\end{proposition}
{\bf Proof.} Let $W=(W_1,W_2)$ be a smooth function such that $L(x,D)W=0\quad\mbox{in}\,\,\Omega.$ We set $E_3=W_1$ and $H_3=W_2.$ Then we introduce functions $H_1,H_2,E_1,E_2$ by formulae (\ref{granit2}). These formulae are equivalent to (\ref{golubka1}) and (\ref{golubka2}). Then the formulae (\ref{oh1}) and (\ref{oh2}) holds true. Hence we may write the equations $L_{(1)}(x,D)W=0$ in form (\ref{10}) and equation $L_{(2)}(x,D)W=0$ in form (\ref{11}). By  (\ref{granit2}) we have first two formulae in (\ref{8}) and (\ref{9}). Finally from first two formulae in (\ref{9})  and (\ref{11}) we obtain the third formula in (\ref{8}). And from first two formulae in (\ref{8})  and (\ref{10}) we obtain the third formula in (\ref{9}). Let $\mbox{\bf f}=(\mbox{\bf f}_1,\mbox{\bf f}_2)$ be a smooth function, $ \mbox{supp}\, \mbox{\bf f}\subset\tilde\Gamma$ such that there exists a solution to the boundary value problem
$$
L(x,D)W=0\quad\mbox{in}\,\,\Omega, \quad \mathcal B(x,D)W=\mbox{\bf f}.
$$
We set $f=(\nu_2\mbox{\bf f}_1, -\nu_1\mbox{\bf f}_1,-\frac{1}{i\omega\gamma}
\mbox{\bf f}_2+\frac{h}{\omega^2\gamma\mu g}\partial_{\vec\tau}\mbox{\bf f}_1).
$ Then if the functions $H$ and $E$ are defined as above we have
$\nu\times E\vert_{\partial\Omega}=f $ and  $ \mbox{supp}\, f\subset\tilde\Gamma .$ Therefore $\Lambda_{\mu,\gamma} f$ is given. The formula (\ref{2}) implies that $H_3\vert_{\tilde \Gamma}=W_2$ is given and $(\nu_1H_2-\nu_2 H_1)\vert_{\tilde \Gamma}$ is given.  Using the formula
$$
(\nu_1H_2-\nu_2 H_1)=-\frac{1}{i\omega\mu }\partial_\nu E_3+\frac{h}{i\omega\mu }(\vec\nu,E')=-\frac{1}{i\omega\mu }\partial_\nu E_3+\frac{h}{\omega^2\gamma\mu }\partial_{\vec\tau} H_3+\frac{h}{\omega^2\gamma\mu }(\nu_1H_2-\nu_2 H_1)\quad \mbox{on}\,\,\tilde \Gamma$$
we obtain that $\partial_\nu E_3=\partial_\nu W_1$ is given on $\tilde \Gamma.$ $\blacksquare$
\bigskip

\section{\bf Step 2: Construction of the operators $P_B$ and $T_B$.}

Let $B$ be a $2\times 2$ matrix with elements from $C^{5+\alpha}
(\overline\Omega)$ with $\alpha\in (0,1)$ and $\widehat x$ be some fixed
point from $\Omega.$
By Proposition 9 of \cite{IY} for the equation
\begin{equation}\label{nina}
(2\partial_{\overline z}+B)u=0\quad
\mbox{in }\,\,\Omega,
\end{equation}
we  can construct solutions $U_{0,k}$ such that
$$
U_{0,k}(\hat x)=\vec e_k,\quad \forall k\in\{1,2\}.
$$
Consider the matrix
$$\Pi (x)=(U_{0,1}(x),U_{0,2}(x)).
$$
Then
$$
\left(\partial_{\overline z}+\frac 12\mbox{tr} B\right) \mbox{det} \Pi=0
\quad \mbox{in}\,\,\Omega.
$$
Hence there exists a holomorphic function $q(z)$ such that
$\mbox{det}\,\Pi=q(z)e^{- \frac 12 \partial^{-1}_{\bar z}(\mbox{tr}\, B)}$
(see \cite{VE}).
By $\mathcal Q$ we denote the set of zeros of the function $q$ on $\overline
\Omega$ :  $\mathcal Q=\{z\in \overline\Omega; \, q(z)=0\}.$
Obviously $card\, \mathcal Q<\infty.$ By $\kappa$ we denote the highest order
of zeros of the function $q$ on $\overline \Omega.$

Using  Proposition 9 of \cite{IY}
we construct solutions $\widetilde U_{0,k}$ to problem (\ref{nina}) such that
$$
\widetilde U_{0,k}( x)=\vec e_k\quad k\in\{1,2\}\quad
\forall x\in \mathcal Q.
$$
Set $\widetilde \Pi (x)=(\widetilde U_{0,1},\widetilde U_{0,2}).$ Then there exists a holomorphic function $\widetilde q$
such that $\mbox{det}\,\widetilde \Pi=\widetilde q(z)e^{-\frac 12
\partial^{-1}_{\bar z}(\mbox{tr}\, B)}.$
Let $\widetilde{\mathcal Q}=\{z\in \overline\Omega; \thinspace
\widetilde q(z)=0\}$ and $\widetilde \kappa$ the highest order of zeros of
the function $\widetilde q.$

By $\widetilde U_{0,k}(x) = \vec{e_k}$ for $x \in \mathcal{Q}$, we see that
$$
\widetilde{\mathcal Q}\cap\mathcal Q=\emptyset.
$$
Therefore there exists a holomorphic function $r(z)$ such that

$$r\vert_{\mathcal Q}=0\quad\mbox{and}\quad (1-r(z))
\vert_{\widetilde{\mathcal Q}}=0
$$
and the orders of zeros of the function $r$ on $\mathcal Q$ and the function
$1-r(z)$ on $\widetilde {\mathcal Q}$ are greater than or equal to the
$\max\{\kappa,\widetilde \kappa\}.$

We set
\begin{equation}\label{victory}
P_{B}f=\frac 12 \Pi\partial^{-1}_{\overline z} (\Pi^{-1}rf)+\frac 12
\widetilde \Pi\partial^{-1}_{\overline z} (\widetilde \Pi^{-1}(1-r)f).
\end{equation}
Then
$$
P_{B}^*f=-\frac 12 r (\Pi^{-1})^*\partial^{-1}_{\overline z} (\Pi^*f)
-\frac 12(1-r) (\widetilde \Pi^{-1})^*\partial^{-1}_{\overline z}
(\widetilde \Pi^*f).
$$

We have
\begin{proposition}\label{gik}
The linear operators $P_{B}, P_B^*\in {\mathcal L} (L^2(\Omega),
W^1_2(\Omega))$ solve the differential equations
\begin{equation}\label{oblom11}
(-2\partial_{\overline z}+B^*)P^*_{B}g=g, \quad
(2\partial_{\overline z}+B)P_{B}g=g\quad \mbox{in}\,\,\Omega.
\end{equation}
\end{proposition}
{\bf Proof.}
Since
$\partial_{\overline z}\Pi=-\frac 12 B\Pi$
and $\partial_{\overline z}\widetilde\Pi=-\frac 12 B\widetilde\Pi$,
short computations imply
$$
\partial_{\overline z}P_{B}f=\partial_{\overline z}\{\frac 12
\Pi\partial^{-1}_{\overline z} (\Pi^{-1}rf)+\frac 12
\widetilde \Pi\partial^{-1}_{\overline z} (\widetilde \Pi^{-1}(1-r)f)\}
$$
$$
= \frac 12(\partial_{\overline z} \Pi)\partial^{-1}_{\overline z}
(\Pi^{-1}rf)+\frac 12 (\partial_{\overline z}\widetilde \Pi)
\partial^{-1}_{\overline z} (\widetilde \Pi^{-1}(1-r)f)\}
$$
$$
+ \frac 12 \Pi (\Pi^{-1}rf)
+ \frac 12 \widetilde \Pi (\widetilde \Pi^{-1}(1-r)f)
$$
$$
= -\frac 14 B\Pi\partial^{-1}_{\overline z} (\Pi^{-1}rf)
- \frac 14 B\widetilde \Pi\partial^{-1}_{\overline z}
(\widetilde \Pi^{-1}(1-r)f)
$$
$$
+ \frac 12 rf+\frac 12 (1-r)f=-\frac 12 BP_{B}f+\frac 12 f.
$$
Hence the second equality in (\ref{oblom11}) is proved. In order to prove
the first one observe that
 since
$
\Pi \Pi^{-1}=E
$ on $\overline\Omega\setminus \mathcal Q$. The differentiation of this
identity gives
$$
0=\partial_{\overline z} (\Pi\Pi^{-1})=\partial_{\overline z}
\Pi\Pi^{-1}+\Pi\partial_{\overline z}\Pi^{-1}=-\frac 12 B\Pi\Pi^{-1}
+\Pi\partial_{\overline z}\Pi^{-1}.
$$
This equality can be written as
$
\Pi\partial_{\overline z}\Pi^{-1}=\frac 12 B.
$
Multiplying both sides of this equality by $ \Pi^{-1}$ we have
$$
\partial_{\overline z}\Pi^{-1}=\frac 1 2 \Pi^{-1} B\quad \mbox{on}\quad
\overline\Omega\setminus \mathcal Q.
$$
Next we take the adjoint for the left- and the right-hand sides of the above
equality:
$$
 (\partial_{\overline z}\Pi^{-1})^*
= \partial_{\overline z}(\Pi^{-1})^*=(\frac 1 2 \Pi^{-1} B)^*
=\frac 12 B^*(\Pi^{-1})^*\quad \mbox{on}\quad\overline\Omega\setminus
\mathcal Q.
$$
Observing that $(\Pi^{-1})^*=(\Pi^*)^{-1}$, we obtain
\begin{equation}\label{XZ}
 \partial_{\overline z}(\Pi^*)^{-1}=\frac 12 B^*(\Pi^*)^{-1}\quad
\mbox{on}\quad\overline\Omega\setminus \mathcal Q.
\end{equation}
 Similarly we obtain
 \begin{equation}\label{ZX}
\partial_{\overline z}(\widetilde\Pi^*)^{-1}
=\frac 12 B^*(\widetilde \Pi^*)^{-1}\quad \mbox{on}\quad
\overline\Omega\setminus\widetilde{ \mathcal Q}.
\end{equation}
Denote by $\Pi_{ij}$ the cofactor of the $i,j$-element of the matrix $\Pi$ and by $\tilde \Pi_{ij}$ the cofactor of the $i,j$-element of the matrix $\tilde \Pi.$ Setting
$$
\Gamma=e^{\frac 12 \partial^{-1}_{\bar z}(\mbox{tr}\, B)}\left(\begin{matrix}
\Pi_{11} &\Pi_{21} \\
 \Pi_{12} &\Pi_{22} \end{matrix}\right),\quad \widetilde\Gamma
= e^{\frac 12 \partial^{-1}_{\bar z}(\mbox{tr}\, B)}\left(\begin{matrix}
\widetilde \Pi_{11} &\widetilde \Pi_{21} \\
 \widetilde \Pi_{12} &\widetilde \Pi_{22} \end{matrix}
\right),
$$
we can write the matrices $(\Pi^*)^{-1},(\widetilde\Pi^*)^{-1}$ as
$$
 (\Pi^*)^{-1}=\frac{1}{q(z)}\Gamma^*\quad \mbox{on}\quad
\overline\Omega\setminus \mathcal Q,\quad (\widetilde\Pi^*)^{-1}
=\frac{1}{\widetilde q(z)}\widetilde\Gamma^*\quad \mbox{on}\quad
\overline\Omega\setminus\tilde{ \mathcal Q}.
$$
Then (\ref{XZ}) and (\ref{ZX}) imply
$$
\partial_{\overline z}(\Gamma^*)^{-1}-\frac 12 B^*(\Gamma^*)^{-1}=
\partial_{\overline z}(\widetilde\Gamma^*)^{-1}
-\frac 12 B^*(\widetilde \Gamma^*)^{-1}=0\quad \mbox{on}\quad\overline\Omega.
$$
Since $r/q$ and $(1-r)/\widetilde q$ are smooth functions,
the above equalities yield
\begin{equation}\label{ZZZ}
\partial_{\overline z}(r(z)(\Pi^*)^{-1})
= \frac {r(z)}{2} B^*(\Pi^*)^{-1}, \quad
\partial_{\overline z}((1-r(z))(\widetilde\Pi^*)^{-1})
= \frac {1-r(z)}{2} B^*(\widetilde \Pi^*)^{-1}\quad \mbox{in}\quad\Omega.
\end{equation}
Using (\ref{ZZZ}), we compute
$$
\partial_{\overline z}P_{B}^*f=-\partial_{\overline z}\{\frac 12 r(z)
(\Pi^{-1})^*\partial^{-1}_{\overline z} (\Pi^*f)+\frac 12(1-r(z))
(\widetilde \Pi^{-1})^*\partial^{-1}_{\overline z} (\widetilde \Pi^*f)\}
$$
$$
= -\frac 12 \partial_{\overline z}( r(z)(\Pi^{-1})^*)\partial^{-1}
_{\overline z} (\Pi^*f)-\frac 12 \partial_{\overline z}
( (1-r(z))(\widetilde\Pi^{-1})^*)\partial^{-1}_{\overline z}
(\widetilde \Pi^*f)
$$
$$
-\frac 12  r(z)(\Pi^{-1})^* \Pi^*f-\frac 12 (1-r(z))(\widetilde\Pi^{-1})^*
\widetilde\Pi^*f
$$
$$
= -\frac 14  r(z)B^*(\Pi^{-1})^*\partial^{-1}_{\overline z} (\Pi^*f)-\frac 14
(1-r(z))B^*(\widetilde\Pi^{-1})^*\partial^{-1}_{\overline z}
(\widetilde \Pi^*f)
$$
$$
-\frac 12  r(z)f-\frac 12 (1-r(z))f=\frac 12 B^*P_{B}^*f-\frac 12 f.
$$
The proof of Proposition \ref{gik} is complete. $\blacksquare$

In a similar way we construct matrices $\Pi_0, \widetilde \Pi_0$, an
antiholomorphic function $r_0(\overline z)$ and operators
\begin{equation}\label{giorgi}
T_{B}f=\frac 12 \Pi_0\partial^{-1}_{z} (\Pi_0^{-1} r_0(\overline z)f)
+ \frac 12 \widetilde \Pi_0\partial^{-1}_{ z}
(\widetilde \Pi_0^{-1}(1-r_0(\overline z))f)
\end{equation}
and
\begin{equation}\label{giorgi1}
T_{B}^*f=-\frac 12  r_0(\overline z) (\Pi_0^{-1})^*\partial^{-1}_{ z}
(\Pi_0^*f)-\frac 12(1-r_0(\overline z)) (\widetilde \Pi_0^{-1})^*
\partial^{-1}_{z} (\widetilde \Pi_0^*f).
\end{equation}

For any matrix $B\in C^{5+\alpha}(\overline \Omega), \alpha\in (0,1)$,  the linear
operators $T_{B}$ and $T^*_{B}$ solve the differential equation
\begin{equation}\label{oblom1}
(2\partial_z+B)T_{B}g=g\quad \mbox{in}\,\,\Omega; \quad
(-2\partial_{ z}+B^*)T^*_{B}g=g\quad \mbox{in}\,\,\Omega.
\end{equation}

Next we introduce two operators
\begin{equation}\label{NaNa}
\widetilde {\mathcal R}_{\tau, B}g = e^{\tau(\overline\Phi-\Phi)}T_B
(e^{\tau(\Phi-\overline\Phi)}g), \quad
{\mathcal R}_{\tau, B}g = e^{\tau(\Phi-\overline\Phi)}
P_B(e^{\tau(\overline\Phi-\Phi)}g).
\end{equation}

\section{\bf Step 3: Construction of complex geometric optics solutions.}

In this step, we will construct two complex geometric optics solutions
$u_1$ and $v$ respectively for operators $L_1(x,D)$ and $L_2(x,D).$

%
%

As the phase function for such a solution  we consider a holomorphic
function $\Phi(z)$ such that $$ \Phi(z)
=\varphi(x_1,x_2)+i\psi(x_1,x_2) $$  with real-valued $\varphi$ and
$\psi.$ For some $\alpha\in (0,1)$ the function $\Phi$ belongs to
$C^{6+\alpha}(\overline{\Omega}).$ Moreover
\begin{equation}\label{zzz}
\partial_{\bar z}\Phi = 0 \quad \mbox{in}
\,\,\Omega, \quad\mbox{Im}\,\Phi\vert_{\Gamma_0}=0.
\end{equation}
 Denote by $\mathcal H$ the set
of all the critical points of the function $\Phi$:
$$
\mathcal H = \{z\in\overline\Omega; \thinspace
\Phi' (z)=0\}.
$$
Assume that $\Phi$ has no critical points on
$\overline{\widetilde\Gamma}$, and that all critical points  are
nondegenerate:
\begin{equation}\label{mika}
\mathcal H\cap \partial\Omega=\emptyset,\quad
\Phi'' (z)\ne 0, \quad \forall z\in
\mathcal H.
\end{equation}

The following proposition
asserts the convergence and was proved
in \cite{IUY}.

\begin{proposition}\label{Proposition -1}
Let $\widetilde x$ be an arbitrary point in the simply connected domain $\Omega.$ There exists a
sequence of functions $\{\Phi_\epsilon\}_{\epsilon\in(0,1)}$
satisfying (\ref{zzz}), (\ref{mika}) and there exists a sequence
$\{\widetilde x_\epsilon\}, \epsilon\in (0,1)$ such that
\begin{equation}\label{bobik1}
\widetilde x_\epsilon \in \mathcal H_\epsilon
= \{z\in\overline\Omega; \thinspace
\Phi_\epsilon'(z)=0 \},\quad \widetilde x_\epsilon\rightarrow \widetilde
x\,\,\mbox{ as}\,\, \epsilon\rightarrow +0.
\end{equation}
and
\begin{equation}\label{bobik2}
\mbox{Im}\,\Phi_\epsilon(\widetilde x_\epsilon)\notin \{\mbox{Im}\,
\Phi_\epsilon(x); \thinspace x\in \mathcal H_\epsilon\setminus
\{\widetilde{x_\epsilon}\}\} \,\,\mbox{and}
\,\,\mbox{Im}\,\Phi_\epsilon(\widetilde x_\epsilon) \ne 0.
\end{equation}
\end{proposition}

Let the function $\Phi$ satisfy (\ref{zzz}), (\ref{mika}) and $\widetilde x$
be some point from $\mathcal H.$
Without loss of generality, we may assume that $\widetilde \Gamma$ is an arc
with the endpoints $x_\pm.$

Consider the following operator:
\begin{eqnarray}\label{ooo} L_1(x,D)=4\partial_z\partial_{\overline z}+2
A_1\partial_z+2B_1\partial_{\overline z}+Q_1 \nonumber\\
= (2\partial_z+B_1)(2\partial_{\overline z}+A_1)+Q_1(1)
= (2\partial_{\overline z}+A_1)(2\partial_z+B_1)+Q_1(2) .
\end{eqnarray}
Here
$$
Q_1(1)=-2\partial_z A_1-B_1A_1+Q_1,\quad Q_2(1)=-2\partial_{\overline z}
B_1-A_1B_1+Q_1.
$$

Let $U_{0}=(U_{0,1}, U_{0,2}),\widetilde U_{0}=(\tilde U_{0,1}, \tilde U_{0,2})\in C^{6+\alpha}(\overline \Omega)$
be a nontrivial solution to the boundary value problem:
\begin{equation}\label{-55!}
\mathcal K(x,D)(U_{0},\widetilde U_{0})=(2\partial_{\overline z}U_{0} +A_1
U_{0}, 2\partial_{ z}\widetilde U_{0} +B_1 \widetilde
U_{0})=0\quad\mbox{in}\,\,\Omega,\end{equation}
\begin{equation}\label{-55!!} U_{0}+\mbox{\bf I}\widetilde U_{0}=0\quad
\mbox{on}\,\,\Gamma_0,
\end{equation}
where
$$
\mbox{\bf I}=\left ( \begin{matrix} 1 &0\\0&-1  \end{matrix}\right ).
$$
We have
\begin{proposition}\label{nikita}
Let $A_1,B_1\in C^{5+\alpha}(\bar \Omega)$ for some $\alpha\in (0,1),$
$\vec r_{0,k},\dots ,\vec r_{2,k}\in \Bbb C^3$ be arbitrary vectors
and $x_1,\dots, x_k$ be mutually distinct arbitrary  points from the domain $\Omega.$
There exists a solution $(U_{0},\widetilde U_{0})\in C^{6+\alpha}(\overline \Omega)$ to
problem (\ref{-55!}), (\ref{-55!!})  such that
\begin{equation}\label{xoxo1}
\partial_z^j U_{0}(x_k)=\vec r_{j,k}\quad
\forall k\in \{1,\dots, 5\},
\end{equation}
\begin{equation}\label{xoxo1u}
\lim_{x\rightarrow x_\pm}\frac{ \vert U_0(x)\vert}{\vert x-x_\pm\vert^{98}}
=\lim_{x\rightarrow x_\pm}\frac{ \vert \widetilde U_0(x)\vert}
{\vert x-x_\pm\vert^{98}}=0.
\end{equation}
\end{proposition}
In (\ref{xoxo1u}), the number $98$ has no special sense.
As exponent we can choose
sufficiently large $m \in \N$ and henceforth we choose such a number
$98$.

{\bf Proof.} Let us fix a point $\widetilde x$ from $\mathcal
H\setminus \{\widetilde x\} .$ By Proposition 4.2 of \cite{IUY3}
there exists a holomorphic function $a(z)\in C^7(\overline\Omega)$
such that $Im\, a\vert_{\Gamma_0}=0,$ $a(\widetilde x)=1$ and $a$
vanishes at each point of the set $\{x_\pm\}\cup\mathcal H\setminus
\{\widetilde x\}$. Let $(U_{0,0} ,\widetilde U_{0,0})\in
C^{6+\alpha}(\overline\Omega)$ be a solution to problem (\ref{-55!})
such that $U_{0,0}(\widetilde x)=\vec z.$ Since $(U_0, \widetilde
U_0)=(a^{100}U_{0,0},\overline a^{100} \widetilde U_{0,0})$ solves
equations (\ref{-55!}) and satisfies (\ref{-55!!}), the
proof of the proposition is completed. $\blacksquare$

Now we start the construction of complex geometric optics solution.
Let the pair $(U_0,\widetilde U_0)$ be defined by Proposition
\ref{nikita}.  Short computations and (\ref{ooo})  yield
\begin{equation}\label{oi}
L_1(x,D) (U_0e^{\tau\Phi})=Q_1(1)U_0e^{\tau\Phi},\quad L_1(x,D) (\widetilde
U_0e^{\tau\overline \Phi})=Q_1(2)\widetilde U_0 e^{\tau\overline\Phi}. \end{equation}

Let $e_1,e_2$ be smooth functions such that
\begin{equation}\label{short}
\mbox{supp}\,e_1\subset\subset\mbox{supp}\,e=1,\quad e_1+e_2=1\quad
\mbox{on}\,\,\Omega,\quad
\end{equation}
and $e_1$ vanishes in a neighborhood of $\partial\Omega$ and $e_2$
vanishes in a neighborhood of the set $\mathcal H.$

For any positive $\epsilon$ denote $G_\epsilon=\{x\in \Omega; \thinspace
dist (\mbox{supp}\,e_1,x)>\epsilon\}.$ We have
\begin{proposition} \label{popo}
Let $\overline G_\epsilon\cap \mbox{supp}\,e =\emptyset$, $B, q\in C^{5+\alpha}(\overline\Omega)$  for some positive
$\alpha\in (0,1)$  and $\widetilde q\in W^1_p(\overline\Omega)$ for some
$p>2.$ Suppose that $q\vert_{\mathcal H}=\widetilde q\vert_{\mathcal
H}=0.$  There exist  functions $m_{\pm,\tilde x}\in C^2(\overline
G_\epsilon), \tilde x\in \mathcal H$  independent of $\tau$ such that the
asymptotic formulae hold true:
\begin{eqnarray}\label{50}
\widetilde{\mathcal R}_{\tau, B}(e_1(q+\frac{\widetilde q}{\tau}))\vert_{\overline
G_\epsilon}
= e^{\tau(\overline\Phi-\Phi)}\left (\sum_{\tilde x\in\mathcal H}\frac{m_{+,\tilde x} e^{2i\tau\psi
(\widetilde x)}}{\tau^2}+o_{C^2(\overline
G_\epsilon)}(\frac{1}{\tau^2})\right)
\quad\mbox{as}\,\vert\tau\vert\rightarrow +\infty ,\\
\quad{\mathcal R}_{\tau, B} (e_1(q + \frac{\widetilde q}{\tau}))\vert_{\overline
G_\epsilon}
=e^{\tau(\Phi-\overline\Phi)}\left (\sum_{\tilde x\in\mathcal H}\frac{m_{-,\tilde x}
e^{-2i\tau\psi(\widetilde x)}}{\tau^2} +o_{C^2(\overline
G_\epsilon)}(\frac{1}{\tau^2})\right
)\quad\mbox{as}\,\vert\tau\vert\rightarrow +\infty.
\end{eqnarray}
\end{proposition}

Denote \begin{equation}\label{soika} q_1=P_{A_1}(Q_1(1)U_{0})-M_1,\quad q_2=
T_{B_1}(Q_1(2)\widetilde
U_{0})-M_2\in C^{5+\alpha}(\bar \Omega),\end{equation}
 where the functions
$M_1\in Ker (2\partial_{\overline z}+A_1)$ and $M_2\in Ker
(2\partial_z+B_1) $ are taken such that
\begin{equation}\label{kl}
q_1(x)=q_2(x)=0, \quad
\forall x\in \mathcal H.
\end{equation}
Moreover by (\ref{xoxo1u}) we can assume that
\begin{equation}\label{bin1}
\lim_{x\rightarrow x_\pm}\frac{ \vert q_1(x)\vert}{\vert x-x_\pm\vert^{98}}
=\lim_{x\rightarrow x_\pm}\frac{ \vert q_2(x)\vert}
{\vert x-x_\pm\vert^{98}}=0.
\end{equation}

Next we introduce the functions $U_{-1},\widetilde U_{-1}$ as a solutions to the following
boundary value problems:
\begin{equation}\label{zad-1}
\mathcal K(x,D)(U_{-1},\widetilde U_{-1})=0\quad\mbox{in}\,\,\Omega,\\\quad
(U_{-1}+\mbox{\bf I}\widetilde
U_{-1})\vert_{\Gamma_0}=\mbox{\bf I}(\frac{q_{1}}{2{\Phi '}}
+ \frac{q_{2}}{2{\bar\Phi '}})+(0, -\frac{1}{\partial_\nu\psi}\partial_\nu( U_{0}+\tilde U_{0}),\vec e_2)).
\end{equation}

We set $p_1=-Q_1(2)(\frac{e_1q_1}{2{\Phi '}}-U_{-1})+L_1(x,D)
(\frac{e_2q_1}{2{\Phi '}})$, $p_2=-Q_1(1)(\frac{e_1q_2}{2{\bar\Phi '}}-\widetilde
U_{-1})+L_1(x,D) (\frac{e_2q_2}{2{\bar\Phi '}}),$
$\widetilde q_2=T_{B_1}p_2-\widetilde M_2, \widetilde q_1=P_{A_1}p_1
-\widetilde M_1$, where  $\widetilde M_1\in Ker (2\partial_{\overline z}+A_1)$
and $\widetilde M_2\in Ker (2\partial_z+B_1)$ are taken such that
\begin{equation}\label{gandon1}
\widetilde q_1( x)
=\widetilde q_2(x)=0,\quad \forall x\in\mathcal H.
\end{equation}

By Proposition \ref{popo}, there exist functions $m_{\pm,\tilde x}\in
C^2(\overline
G_\epsilon)$ such that
\begin{equation}\label{50l}
\widetilde{\mathcal R}_{\tau, B_1}(e_1(q_1+\frac{\widetilde
q_1}{\tau})) \vert_{\overline
G_\epsilon}=
e^{\tau(\overline\Phi-\Phi)}\left (\sum_{\tilde x\in\mathcal H}\frac{m_{+,\tilde x} e^{2i\tau\psi
(\widetilde
x)}}{\tau^2}+o_{W^1_2(\overline
G_\epsilon)}(\frac{1}{\tau^2})\right)
\quad\mbox{as}\,\vert\tau\vert\rightarrow +\infty
\end{equation}
and
\begin{equation}\label{50ll}
\quad{\mathcal R}_{\tau, A_1} (e_1(q_2 + \frac{\widetilde
q_2}{\tau})) \vert_{\overline
G_\epsilon} =e^{\tau(\Phi-\overline\Phi)}
\left (\sum_{\tilde x\in\mathcal H}\frac{m_{-,\tilde x} e^{-2i\tau\psi(\widetilde x)}}{\tau^2}
+o_{W^1_2(\overline
G_\epsilon)}(\frac{1}{\tau^2})\right
)\quad\mbox{as}\,\vert\tau\vert\rightarrow +\infty.
\end{equation}

For any $\tilde x\in \mathcal H$ we introduce the functions ,
$a_{\pm,\tilde x},b_{\pm,\tilde x}\in C^2(\overline \Omega)$ as solutions to the boundary value problem
\begin{equation}\label{lobster}
\mathcal K(x,D)(a_{\pm,\tilde x},b_{\pm,\tilde x})=0\quad\mbox{in}\,\,\Omega,\quad
(a_{\pm,\tilde x}+\mbox{\bf I}b_{\pm,\tilde x})\vert_{\Gamma_0}=\pm m_{\pm,\tilde x}.
\end{equation}

Since by (\ref{gandon1}) the functions $\frac{\widetilde q_1}{2{\Phi '}},\frac{\widetilde q_2}
{2{\bar\Phi '}}$ belong to the space $ W^1_2(\partial\Omega)$
 there exists a solution $(U_{-2},\widetilde U_{-2})
\in W^1_2(\overline\Omega)$ to the
boundary value problem
\begin{eqnarray}\label{zad-11} \mathcal K(x,D)(U_{-2},\widetilde U_{-2})
=0\quad\mbox{in}\,\,\Omega,\\
(U_{-2}+\mbox{\bf I}\widetilde U_{-2})\vert_{\Gamma_0}=((\frac{\widetilde
q_1}{2{\Phi '}}+\frac{\widetilde q_2}{2\bar\Phi'},\vec e_1), (-\frac{\widetilde
q_1}{2{\Phi '}}+\frac{\widetilde q_2}{2
\overline\Phi'},\vec e_2))\nonumber\\-\frac{1}{i\partial_\nu\psi}(0, \partial_\nu(U_{-1}-\frac{e_2q_1}{2\Phi '}+\tilde U_{-1}-\frac{e_2q_2}{2\bar \Phi'},\vec e_2)).\nonumber
\end{eqnarray}

We introduce the functions $U_{0,\tau}, \widetilde U_{0,\tau}
\in W^1_2(\Omega)$ by
\begin{equation}\label{zad1}
U_{0,\tau}=U_0+\frac{U_{-1}-e_2q_1/2\Phi'}{\tau}+\frac{1}{\tau^2}(\sum_{\tilde x\in\mathcal H}( e^{2i\tau\psi(\widetilde
x)}a_{+,\tilde x}+e^{-2i\tau\psi(\widetilde x)}a_{-,\tilde x})+U_{-2}-\frac{\widetilde
q_1 e_2}{2{\Phi '}})
\end{equation}
and
\begin{equation}\label{zad2}
\widetilde U_{0,\tau}=\widetilde U_0+\frac{\widetilde U_{-1}-e_2
q_2/2{\bar\Phi '}}{\tau}+\frac{1}{\tau^2}(\sum_{\tilde x\in\mathcal H}(
e^{2i\tau\psi(\widetilde x)}b_{+,\tilde x}+e^{-2i\tau\psi(\widetilde
x)}b_{-,\tilde x})+\widetilde U_{-2}-\frac{\widetilde q_2 e_2}{2
\overline\Phi'}).
\end{equation}

Simple computations and Proposition 8 of \cite{IY} imply  for any $p\in
(1,\infty)$ the asymptotic formula:
\begin{eqnarray}\label{251}
L_1(x,D)(-e^{\tau\Phi}\widetilde{\mathcal R}_{\tau, B_1}(e_1(q_1+\widetilde
q_1/\tau))-\frac{e_2(q_1+\widetilde
q_1/\tau)e^{\tau\Phi}}{2\tau{\Phi '}}-e^{\tau\overline\Phi}{\mathcal
R}_{\tau,
A_1}(e_1(q_2+\widetilde q_2/\tau))\nonumber\\-\frac{e_2(q_2+\widetilde q_2/\tau)e^{\tau\overline\Phi}}{2\tau{\bar\Phi '}})
                                                      \nonumber
= -L_1(x,D)(e^{\tau\Phi}\widetilde{\mathcal R}_{\tau, B_1}
(e_1(q_1+\widetilde q_1/\tau))+\frac{e_2(q_1+\widetilde
q_1/\tau)e^{\tau\Phi}}{2\tau{\Phi '}})\\
- L_1(x,D)(e^{\tau\overline\Phi}{\mathcal R}_{\tau,
A_1}(e_1(q_2+\widetilde q_2/\tau))+\frac{e_2(q_2+\widetilde q_2/\tau)
e^{\tau\overline\Phi}}{2\tau{\bar\Phi '}})\nonumber\\
= -Q_1(2)e^{\tau\Phi}\widetilde {\mathcal
R}_{\tau,B_1}(e_1(q_1+\widetilde q_1/\tau))-Q_1(1)e^{\tau\overline\Phi}{\mathcal
R}_{\tau,A_1}(e_1(q_2+\widetilde q_2/\tau))\nonumber\\
-e^{\tau\Phi}L_1(x,D) (\frac{e_2(q_1+\widetilde
q_1/\tau)}{2\tau{\Phi '}})-e^{\tau\overline\Phi}L_1(x,D)
(\frac{e_2(q_2+\widetilde q_2/\tau)}{2\tau{\bar\Phi '}})
\nonumber\\-Q_1(2)\widetilde U_0
e^{\tau\overline\Phi}-Q_1(1)U_0e^{\tau\Phi}
+ \frac 1\tau(Q_1(2)(\frac{e_1q_1}{2{\Phi '}}-U_1)+L_1(x,D)
(\frac{e_2q_1}{2{\Phi '}}))e^{\tau\Phi} \nonumber\\
+ \frac 1\tau(Q_1(1)(\frac{e_2q_2}{2\overline\Phi'}-\tilde U_1)+L_1(x,D) (\frac{e_2q_2}{2\overline\Phi'}))e^{\tau\overline\Phi}
= - \frac{1}{\tau}Q_1(2) U_{-1}e^{\tau\Phi}-\frac{1}{\tau}Q_1(1)\widetilde
U_{-1}e^{\tau\overline\Phi}\nonumber\\
- Q_1(2)\widetilde U_0 e^{\tau\overline\Phi}-Q_1(1)U_0e^{\tau\Phi}
+e^{\tau\varphi}o_{L^p(\Omega)}(\frac{1}{\tau}).
\end{eqnarray}
We set
$$
U=U_{0,\tau}e^{\tau\Phi}+U_{0,\tau}e^{-\tau\bar\Phi}-e^{\tau\Phi}\widetilde{\mathcal R}_{\tau, B_1}(e_1(q_1+\widetilde
q_1/\tau))-e^{\tau\overline\Phi}{\mathcal
R}_{\tau,
A_1}(e_1(q_2+\widetilde q_2/\tau)).
$$

Using (\ref{251}) we prove the following proposition.
\begin{proposition}\label{Proposition 00}
For any $p>1$, we have the asymptotic formula:
\begin{equation}\label{249}
L_1(x,D)U=e^{\tau\varphi}o_{L^p(\Omega)}(\frac{1}{\tau}) ,
\end{equation}
\begin{equation}\label{zonzon} \mathcal B(x,D)U\vert_{\Gamma_0}=e^{\tau\varphi}\left(\begin{matrix} o_{W^1_2(\Gamma_0)}(\frac {1}{\tau^2})\\ O_{W^1_2(\Gamma_0)}(\frac {1}{\tau^2})\end{matrix}\right )
.
\end{equation}
\end{proposition}

{\bf Proof.} By (\ref{zzz}), (\ref{50l})-(\ref{lobster})
and (\ref{zad-1})-(\ref{zad2}), we have
$$
(U_{0,\tau}e^{\tau \Phi}+\widetilde U_{0,\tau} e^{\tau \overline
\Phi}-e^{\tau\Phi}\widetilde{\mathcal R}_{\tau, B_1}(e_1(q_1+\widetilde
q_1/\tau))-e^{\tau\overline\Phi}{\mathcal R}_{\tau,
A_1}(e_1(q_2+\widetilde q_2/\tau),\vec e_1)\vert_{\Gamma_0}
$$
$$
= (U_{0,\tau}e^{\tau \varphi}+\widetilde U_{0,\tau} e^{\tau
\varphi}-e^{\tau\varphi}\widetilde{\mathcal R}_{\tau,
B_1}(e_1(q_1+\widetilde q_1/\tau))-e^{\tau\varphi}{\mathcal R}_{\tau,
A_1}(e_1(q_2+\widetilde q_2/\tau)),\vec e_1)\vert_{\Gamma_0}
$$
$$
= e^{\tau\varphi}(U_0+\frac{U_{-1}-e_2q_1/2\Phi'}{\tau}
+\frac{1}{\tau^2}(\sum_{\tilde x\in\mathcal H}( e^{2i\tau\psi(\widetilde
x)}a_{+,\tilde x}+e^{-2i\tau\psi(\widetilde x)}a_{-,\tilde x}) +U_{-2}-\frac{\widetilde q_1
e_2}{2{\Phi '}}) $$
$$
+ \widetilde U_0+\frac{\widetilde U_{-1}-e_2
q_2/2{\bar\Phi '}}{\tau}+\frac{1}{\tau^2}(\sum_{\tilde x\in\mathcal H}(
e^{2i\tau\psi(\widetilde x)}b_{+,\tilde x} +e^{-2i\tau\psi(\widetilde
x)}b_{-,\tilde x})+\widetilde U_{-2} -\frac{\widetilde q_2
e_2}{2{\bar\Phi '}})
$$
$$
-\widetilde{\mathcal R}_{\tau, B_1}(e_1(q_1+\widetilde
q_1/\tau))-{\mathcal R}_{\tau, A_1}(e_1(q_2+\widetilde
q_2/\tau))),\vec e_1)\vert_{\Gamma_0}
$$
$$
= e^{\tau\varphi}(\frac{1}{\tau^2}
\sum_{\tilde x\in\mathcal H}(e^{2i\tau\psi(\widetilde x)}a_{+,\tilde x}+e^{-2i\tau\psi(\widetilde
x)}a_{-,\tilde x} + e^{2i\tau\psi(\widetilde
x)}b_{+,\tilde x}+e^{-2i\tau\psi(\widetilde x)}b_{-,\tilde x})
$$
$$
- \widetilde{\mathcal R}_{\tau, B_1}(e_1(q_1+\widetilde
q_1/\tau))-{\mathcal R}_{\tau, A_1}(e_1(q_2+\widetilde
q_2/\tau)),\vec e_1)\vert_{\Gamma_0}=e^{\tau\varphi}o_{W^1_2(\Gamma_0)}(\frac
{1}{\tau^2}).
$$

The short computations imply
$$
\mathcal I=\partial_{\nu}(U_{0,\tau}e^{\tau \Phi}+\widetilde U_{0,\tau} e^{\tau \overline
\Phi}-e^{\tau\Phi}\widetilde{\mathcal R}_{\tau, B_1}(e_1(q_1+\widetilde
q_1/\tau))-e^{\tau\overline\Phi}{\mathcal R}_{\tau,
A_1}(e_1(q_2+\widetilde q_2/\tau),\vec e_2)\vert_{\Gamma_0}
$$
$$
= (i\tau\partial_{\nu}\psi U_{0,\tau}e^{\tau \varphi}-i\tau\partial_{\nu}\psi \widetilde U_{0,\tau} e^{\tau
\varphi}+
\partial_{\nu}U_{0,\tau}e^{\tau \varphi}+\partial_{\nu} \widetilde U_{0,\tau} e^{\tau
\varphi}
$$
$$
-e^{\tau\varphi}i\partial_{\nu}\psi\widetilde{\mathcal R}_{\tau,
B_1}(e_1(q_1+\widetilde q_1/\tau))+i\partial_{\nu}\psi e^{\tau\varphi}{\mathcal R}_{\tau,
A_1}(e_1(q_2+\widetilde q_2/\tau))
$$
$$-e^{\tau\varphi}\partial_{\nu}\widetilde{\mathcal R}_{\tau,
B_1}(e_1(q_1+\widetilde q_1/\tau))-e^{\tau\varphi}\partial_{\nu}{\mathcal R}_{\tau,
A_1}(e_1(q_2+\widetilde q_2/\tau)),\vec e_2)\vert_{\Gamma_0}
$$
$$
=( e^{\tau\varphi}i\tau\partial_{\nu}\psi(U_0+\frac{U_{-1}-e_2q_1/2\Phi'}{\tau}
+\frac{1}{\tau^2}(\sum_{\tilde x\in\mathcal H}( e^{2i\tau\psi(\widetilde
x)}a_{+,\tilde x}+e^{-2i\tau\psi(\widetilde x)}a_{-,\tilde x} )+U_{-2}-\frac{\widetilde q_1
e_2}{2{\Phi '}})) $$
$$
- e^{\tau\varphi}i\tau\partial_{\nu}\psi( \widetilde U_0+\frac{\widetilde U_{-1}-e_2
q_2/2{\bar\Phi '}}{\tau}+\frac{1}{\tau^2}(\sum_{\tilde x\in\mathcal H}(
e^{2i\tau\psi(\widetilde x)}b_{+,\tilde x} +e^{-2i\tau\psi(\widetilde
x)}b_{-,\tilde x})+\widetilde U_{-2} -\frac{\widetilde q_2
e_2}{2{\bar\Phi '}}))
$$
$$
+( e^{\tau\varphi}\partial_{\nu}(U_0+\frac{U_{-1}-e_2q_1/2\Phi'}{\tau}
+\frac{1}{\tau^2}(\sum_{\tilde x\in\mathcal H}( e^{2i\tau\psi(\widetilde
x)}a_{+,\tilde x}+e^{-2i\tau\psi(\widetilde x)}a_{-,\tilde x}) +U_{-2}-\frac{\widetilde q_1
e_2}{2{\Phi '}})) $$
$$
+e^{\tau\varphi}\partial_{\nu}( \widetilde U_0+\frac{\widetilde U_{-1}-e_2
q_2/2{\bar\Phi '}}{\tau}+\frac{1}{\tau^2}(\sum_{\tilde x\in\mathcal H}(
e^{2i\tau\psi(\widetilde x)}b_{+,\tilde x} +e^{-2i\tau\psi(\widetilde
x)}b_{-,\tilde x})+\widetilde U_{-2} -\frac{\widetilde q_2
e_2}{2{\bar\Phi '}}))
$$

$$
-e^{\tau\varphi}i\tau\partial_{\nu}\psi\widetilde{\mathcal R}_{\tau,
B_1}(e_1(q_1+\widetilde q_1/\tau))+i\tau\partial_{\nu}\psi e^{\tau\varphi}{\mathcal R}_{\tau,
A_1}(e_1(q_2+\widetilde q_2/\tau))
$$
$$-e^{\tau\varphi}\partial_{\nu}\widetilde{\mathcal R}_{\tau,
B_1}(e_1(q_1+\widetilde q_1/\tau))-e^{\tau\varphi}\partial_{\nu}{\mathcal R}_{\tau,
A_1}(e_1(q_2+\widetilde q_2/\tau)),\vec e_2)\vert_{\Gamma_0}.
$$
By (\ref{-55!}) and (\ref{zad-1}) we obtain
$$
\mathcal  I=( e^{\tau\varphi}i\tau\partial_{\nu}\psi(
\frac{1}{\tau^2}(\sum_{\tilde x\in\mathcal H}( e^{2i\tau\psi(\widetilde
x)}a_{+,\tilde x}+e^{-2i\tau\psi(\widetilde x)}a_{-,\tilde x}) +U_{-2}-\frac{\widetilde q_1
e_2}{2{\Phi '}})) $$
$$
- e^{\tau\varphi}i\tau\partial_{\nu}\psi(\frac{1}{\tau^2}(\sum_{\tilde x\in\mathcal H}(
e^{2i\tau\psi(\widetilde x)}b_{+,\tilde x} +e^{-2i\tau\psi(\widetilde
x)}b_{-,\tilde x})+\widetilde U_{-2} -\frac{\widetilde q_2
e_2}{2{\bar\Phi '}}))
$$
$$
+( e^{\tau\varphi}\partial_{\nu}(\frac{U_{-1}-e_2q_1/2\Phi'}{\tau}
+\frac{1}{\tau^2}(\sum_{\tilde x\in\mathcal H}( e^{2i\tau\psi(\widetilde
x)}a_{+,\tilde x}+e^{-2i\tau\psi(\widetilde x)}a_{-,\tilde x}) +U_{-2}-\frac{\widetilde q_1
e_2}{2{\Phi '}})) $$
$$
+e^{\tau\varphi}\partial_{\nu}( \frac{\widetilde U_{-1}-e_2
q_2/2{\bar\Phi '}}{\tau}+\frac{1}{\tau^2}(\sum_{\tilde x\in\mathcal H}(
e^{2i\tau\psi(\widetilde x)}b_{+,\tilde x} +e^{-2i\tau\psi(\widetilde
x)}b_{-,\tilde x})+\widetilde U_{-2} -\frac{\widetilde q_2
e_2}{2{\bar\Phi '}}))
$$

$$
-e^{\tau\varphi}i\tau\partial_{\nu}\psi\widetilde{\mathcal R}_{\tau,
B_1}(e_1(q_1+\widetilde q_1/\tau))+i\tau\partial_{\nu}\psi e^{\tau\varphi}{\mathcal R}_{\tau,
A_1}(e_1(q_2+\widetilde q_2/\tau))
$$
$$-e^{\tau\varphi}\partial_{\nu}\widetilde{\mathcal R}_{\tau,
B_1}(e_1(q_1+\widetilde q_1/\tau))-e^{\tau\varphi}\partial_{\nu}{\mathcal R}_{\tau,
A_1}(e_1(q_2+\widetilde q_2/\tau)),\vec e_2)\vert_{\Gamma_0}.
$$

Using  Proposition \ref{popo}, the formulae (\ref{50l}) and (\ref{50ll})  we obtain
$$
\mathcal I=( e^{\tau\varphi}i\tau\partial_{\nu}\psi(
\frac{1}{\tau^2}(\sum_{\tilde x\in\mathcal H}( e^{2i\tau\psi(\widetilde
x)}a_{+,\tilde x}+e^{-2i\tau\psi(\widetilde x)}a_{-,\tilde x} )+U_{-2}-\frac{\widetilde q_1
e_2}{2{\Phi '}})) $$
$$
- e^{\tau\varphi}i\tau\partial_{\nu}\psi( \frac{1}{\tau^2}(\sum_{\tilde x\in\mathcal H}(
e^{2i\tau\psi(\widetilde x)}b_{+,\tilde x} +e^{-2i\tau\psi(\widetilde
x)}b_{-,\tilde x})+\widetilde U_{-2} -\frac{\widetilde q_2
e_2}{2{\bar\Phi '}}))
$$
$$
+( e^{\tau\varphi}\partial_{\nu}(\frac{U_{-1}-e_2q_1/2\Phi'}{\tau}
+\frac{1}{\tau^2}( \sum_{\tilde x\in\mathcal H}(e^{2i\tau\psi(\widetilde
x)}a_{+,\tilde x}+e^{-2i\tau\psi(\widetilde x)}a_{-,\tilde x}) +U_{-2}-\frac{\widetilde q_1
e_2}{2{\Phi '}})) $$
$$
+e^{\tau\varphi}\partial_{\nu}( \frac{\widetilde U_{-1}-e_2
q_2/2{\bar\Phi '}}{\tau}+\frac{1}{\tau^2}(\sum_{\tilde x\in\mathcal H}(
e^{2i\tau\psi(\widetilde x)}b_{+,\tilde x} +e^{-2i\tau\psi(\widetilde
x)}b_{-,\tilde x})+\widetilde U_{-2} -\frac{\widetilde q_2
e_2}{2{\bar\Phi '}}))
$$

$$
-e^{\tau\varphi}i\partial_{\nu}\psi\sum_{\tilde x\in\mathcal H}\frac{ m_{+,\tilde x} e^{2i\tau\psi(\tilde x)}}{\tau}+e^{\tau\varphi}i\partial_\nu\psi\sum_{\tilde x\in\mathcal H}\frac{ m_{-,\tilde x} e^{-2i\tau\psi(\tilde x)}}{\tau}
$$
$$-e^{\tau\varphi}\sum_{\tilde x\in\mathcal H}\frac{\partial_{\nu} m_{+,\tilde x} e^{2i\tau\psi(\tilde x)}}{\tau^2}-e^{\tau\varphi}\sum_{\tilde x\in\mathcal H}\frac{\partial_{\nu} m_{-,\tilde x} e^{-2i\tau\psi(\tilde x)}}{\tau^2},\vec e_2)\vert_{\Gamma_0}+e^{\tau\varphi}o_{W^1_2(\Gamma_0)}(\frac{1}{\tau}).
$$

Using (\ref{zad-11}) write down $\mathcal I$ as
$$
\mathcal I=( e^{\tau\varphi}i\tau\partial_{\nu}\psi(
\frac{1}{\tau^2}\sum_{\tilde x\in\mathcal H}( e^{2i\tau\psi(\widetilde
x)}a_{+,\tilde x}+e^{-2i\tau\psi(\widetilde x)}a_{-,\tilde x} ) $$
$$
- e^{\tau\varphi}i\tau\partial_{\nu}\psi( \frac{1}{\tau^2}(\sum_{\tilde x\in\mathcal H}(
e^{2i\tau\psi(\widetilde x)}b_{+,\tilde x} +e^{-2i\tau\psi(\widetilde
x)}b_{+,\tilde x}))
$$
$$
+( e^{\tau\varphi}\partial_{\nu}(
\frac{1}{\tau^2}(\sum_{\tilde x\in\mathcal H}( e^{2i\tau\psi(\widetilde
x)}a_{+,\tilde x}+e^{-2i\tau\psi(\widetilde x)}a_{-,\tilde x}) +U_{-2}-\frac{\widetilde q_1
e_2}{2{\Phi '}})) $$
$$
+e^{\tau\varphi}\partial_{\nu}( \frac{1}{\tau^2}(\sum_{\tilde x\in\mathcal H}(
e^{2i\tau\psi(\widetilde x)}b_{+,\tilde x} +e^{-2i\tau\psi(\widetilde
x)}b_{-,\tilde x})+\widetilde U_{-2} -\frac{\widetilde q_2
e_2}{2{\bar\Phi '}}))
$$

$$
-e^{\tau\varphi}i\partial_{\nu}\psi\sum_{\tilde x\in\mathcal H}\frac{ m_{+,\tilde x} e^{2i\tau\psi(\tilde x)}}{\tau}+e^{\tau\varphi}i\partial_\nu\psi\sum_{\tilde x\in\mathcal H}\frac{ m_{-,\tilde x} e^{-2i\tau\psi(\tilde x)}}{\tau}
$$
$$-e^{\tau\varphi}\sum_{\tilde x\in\mathcal H}\frac{\partial_{\nu} m_{+,\tilde x} e^{2i\tau\psi(\tilde x)}}{\tau^2}-e^{\tau\varphi}\sum_{\tilde x\in\mathcal H}\frac{\partial_{\nu} m_{-,\tilde x} e^{-2i\tau\psi(\tilde x)}}{\tau^2},\vec e_2)\vert_{\Gamma_0}+e^{\tau\varphi}o_{W^1_2(\Gamma_0)}(\frac{1}{\tau}).
$$
Finally, applying (\ref{lobster}) we have
$$
\mathcal I=
( e^{\tau\varphi}\partial_{\nu}(
\frac{1}{\tau^2}(\sum_{\tilde x\in\mathcal H}( e^{2i\tau\psi(\widetilde
x)}a_{+,\tilde x}+e^{-2i\tau\psi(\widetilde x)}a_{-,\tilde x}) +U_{-2}-\frac{\widetilde q_1
e_2}{2{\Phi '}}),\vec e_2) $$
$$
+e^{\tau\varphi}\partial_{\nu}( \frac{1}{\tau^2}(\sum_{\tilde x\in\mathcal H}(
e^{2i\tau\psi(\widetilde x)}b_{+,\tilde x} +e^{-2i\tau\psi(\widetilde
x)}b_{-,\tilde x})+\widetilde U_{-2} -\frac{\widetilde q_2
e_2}{2{\bar\Phi '}}),\vec e_2)+e^{\tau\varphi}o_{W^1_2(\Gamma_0)}(\frac{1}{\tau})=e^{\tau\varphi}o_{W^1_2(\Gamma_0)}(\frac{1}{\tau}).
$$
The proof of equality (\ref{zonzon}) is complete now.

 Similarly to (\ref{oi}) we obtain
\begin{eqnarray}\label{250}
L_1(x,D)(U_{0,\tau}e^{\tau\Phi}+\widetilde
U_{0,\tau}e^{\tau\overline\Phi}+\frac{e_2(q_1+\widetilde
q_1/\tau)e^{\tau\Phi}}{2\tau{\Phi '}}+\frac{e_2(q_2+\widetilde
q_2/\tau)e^{\tau\overline\Phi}}{2\tau\overline\Phi'})\nonumber\\
=  Q_1(1)(U_{0,\tau}+\frac{e_2(q_1+\widetilde
q_1/\tau)}{2\tau{\Phi '}})e^{\tau\Phi}
+ Q_1(2)(\widetilde U_{0,\tau}
+\frac{e_2(q_2+\widetilde
q_2/\tau)}{2\tau{\bar\Phi '}})e^{\tau\overline\Phi}.
\end{eqnarray}
By (\ref{250}) and (\ref{251}), we obtain (\ref{249}). $\blacksquare$

Consider the boundary value problem
$$
L(x,{D})u=\Delta u+2A\partial_z u+2B\partial_{\overline z} u+Qu = f
\quad \mbox{in} \quad \Omega, \quad\mathcal B(x,D) u \vert_{\partial \Omega} = 0.
$$
Then we prove a Carleman estimate with boundary terms
whose weight function is degenerate.
\begin{proposition}\label{ETheorem 2.1}
Suppose that $\Phi=\varphi+i\psi$ satisfies (\ref{zzz}),
(\ref{mika}), the coefficients of the operator $L$ matrices $A,B,Q$
belong to $L^\infty(\Omega).$  Then there exist
$\tau_0$ and $C$, independent of $u$ and
$\tau$, such that
\begin{eqnarray}\label{suno4} \vert \tau\vert\Vert
ue^{\tau\varphi}\Vert^2_{L^2(\Omega)}+\Vert
ue^{\tau\varphi}\Vert^2_{W^1_2(\Omega)}+
\tau^2\Vert\vert\frac{\partial\Phi}{\partial z} \vert
ue^{\tau\varphi}\Vert^2_{L^2(\Omega)}\nonumber \\
\le C(\Vert (L(x,D)
u)e^{\tau\varphi}\Vert^2_{L^2(\Omega)}+\vert\tau\vert
\int_{\widetilde\Gamma}(\vert \nabla u\vert^2+\tau^2 u^2)e^{2\tau\varphi}d\sigma)
\end{eqnarray}
for all $\vert\tau\vert>\tau_0$ and all $u\in W^2_2(\Omega), \mathcal B(x,D)u\vert_{\Gamma_0}=0$.
\end{proposition}

 For the scalar equation, the estimate is proved in
\cite{IUY} and \cite{IUY3} for the case of the Dirichelt and Neumann boundary conditions respectively. In order to prove this estimate for the system, it is
sufficient to apply the scalar estimate to each equation in the
system and take an advantage of the second large parameter in order
to absorb the right-hand side.

Using estimate (\ref{suno4}), we obtain
\begin{proposition} \label{vanka}
There exists a constant $\tau_0$ such that for $\vert
\tau\vert\ge \tau_0$ and any $f\in L^2(\Omega)$, there exists a
solution to the boundary value problem
\begin{equation}\label{lola}
L(x,D+i\tau\nabla\varphi)u =f\quad\mbox{in}\,\,\Omega, \quad
\mathcal B(x,D)u\vert_{\Gamma_0}=0
\end{equation}
such that
\begin{equation}\label{2q}
\Vert u\Vert_{W_2^{1,\tau}(\Omega)}/\root\of{\vert\tau\vert}
\le C\Vert  f\Vert_{L^2(\Omega)}.
\end{equation}
Moreover if $f/{\Phi '}\in L^2(\Omega)$, then for any $\vert
\tau\vert\ge \tau_0$  there exists a solution to the boundary value
problem (\ref{lola}) such that
\begin{equation}\label{3q}
\Vert u\Vert_{W_2^{1,\tau}(\Omega)} \le C\Vert
f/{\Phi '}\Vert_{L^2(\Omega)}.
\end{equation}
The constants $C$ in (\ref{2q}) and (\ref{3q}) are independent of
$\tau.$
\end{proposition}

We set ${\mathcal O}_{\epsilon}=\{x\in \Omega; \thinspace dist (x,\partial
\Omega)\le \epsilon\}.$
In order to construct the last term in complex geometric optics solution,
we need the following proposition:

\begin{proposition}\label{Proposition 0}
Let $A,B\in C^{5+\alpha}(\overline\Omega)$ and $Q\in C^{4+\alpha}
(\overline\Omega)$ for some $\alpha\in (0,1),$ $f\in L^p(\Omega)$
for some $p>2$, $dist (\overline\Gamma_0,supp\,f)>0$, $q\in W^1_2(\Gamma_0),$ and $\epsilon$ be a small positive number such that $
\overline{{\mathcal O}_{\epsilon}}\cap{\mathcal
H}=\emptyset.$ Then there exists $C$ independent
of $\tau$ and $\tau_0$ such that for all $\vert\tau\vert>\tau_0$,
there exists a solution $w\in W^1_2(\Omega)$ to the boundary value problem
\begin{equation}\label{mimino}
L(x,D)w=fe^{\tau\Phi}\quad \mbox{in}\,\,\Omega, \quad\mathcal B(x,D)
w\vert_{\Gamma_0} =qe^{\tau\varphi}/\tau
\end{equation}
such that
\begin{equation}\label{mimino111}
\root\of{\vert\tau\vert} \Vert w e^{-\tau\varphi}\Vert_{L^2(\Omega)}
+ \frac{1}{\root\of{\vert\tau\vert} } \Vert (\nabla w)
e^{-\tau\varphi}\Vert_{L^2(\Omega)}+\Vert we^{-\tau\varphi}
\Vert_{H^{1,\tau}(\mathcal O_{\epsilon})} \le C(\Vert
f\Vert_{L^p(\Omega)} + \Vert q \Vert_{W^1_2(\Gamma_0)}).
\end{equation}
\end{proposition}

{\bf Proof.} First let us assume that $f$ is identically equal to
zero. Let $(d,\widetilde d)\in W^1_2(\Omega)\times  W^1_2(\overline\Omega)$
satisfy
\begin{equation}\label{zadnitsa}
\mathcal K(x,D)(d,\widetilde d)=0\quad\mbox{in}\,\Omega, \quad (d+\mbox{\bf I}\widetilde
d)\vert_{\Gamma_0}=q.
\end{equation}
For existence of such a solution see e.g. \cite{Wendland}. By
(\ref{oi}) and (\ref{zadnitsa}), we have
$$
L(x,D)(\frac{d}{\tau}e^{\tau\Phi}+\frac{\widetilde
d}{\tau}e^{\tau\overline\Phi})=\frac{1}{\tau}(Q-2\partial_z A-BA)d
e^{\tau\Phi}+\frac{1}{\tau}(Q-2\partial_{\overline z} B-AB)\widetilde d
e^{\tau\overline\Phi}.
$$
By Proposition \ref{vanka}, there exists  a solution $w$ to the
boundary value problem
$$
L(x,D)\tilde w=-\frac{1}{\tau}(Q-2\partial_z A-BA)d
e^{\tau\Phi}-\frac{1}{\tau}(Q-2\partial_{\overline z}
B-AB)\widetilde d e^{\tau\overline\Phi},\quad{\mathcal B}(x,D)\tilde
w\vert_{\Gamma_0}=0
$$
such that there exists a constant $C>0$ such that
$$
\Vert \tilde we^{-\tau\varphi}\Vert_{H^{1,\tau}(\Omega)}\le \frac{C}
{\root\of{\vert \tau\vert}}\Vert (Q-2\partial_z A-BA)d
e^{i\tau\psi}+(Q-2\partial_{\overline z} B-AB)\widetilde d
e^{-\tau i\psi})\Vert_{L^2(\Omega)}\le \frac{C
}{\root\of{\vert\tau\vert}}\Vert q\Vert_{W^1_2(\Gamma_0)}
$$
for all large $\tau>0$.

Then the function $(\frac{d}{\tau}e^{\tau\Phi}+\frac{\widetilde
d}{\tau}e^{\tau\overline\Phi})+\tilde w$ is a solution to
(\ref{mimino}) which satisfies (\ref{mimino111}) if $f\equiv 0.$

If $f$ is not identically equal zero without the loss of generality we may assume that $q\equiv 0.$ Then we consider the function
$\tilde w=\widetilde e e^{\tau\Phi}\widetilde{\mathcal R}_{\tau,
B}(e_1q_0)$, where $\widetilde e\in
C^\infty_0(\Omega),\,\,\widetilde e\vert_{\mbox{supp}e_1}=1$ and
$q_0=P_{A}f-M,$  where a function $M\in C^5(\bar \Omega)$ belongs to
$Ker\, (2\partial_z+B)$ and chosen such that $q_0\vert_{\mathcal
H}=0.$ Then $ L(x,D)\tilde w=(Q-2\partial_{\overline z}B-AB)\tilde
w+ \widetilde ee_1fe^{\tau\Phi}+2\widetilde
ee^{\tau\Phi}q_0\partial_{\overline
z}e_1+e^{\tau\Phi}(2\partial_{\overline z}+A)(\partial_z
e\widetilde{\mathcal R}_{\tau, B}(e_1q_0)). $ Since, by Proposition
8 of \cite{IY}, the function $\widetilde f(\tau,\cdot)=e^{-\tau\Phi}L(x,D)\tilde
w-f$ can be represented as a sum of two functions, where the first
one equal to zero in a neighborhood of $\mathcal H$ and is bounded
uniformly in $\tau$ in $L^2(\Omega)$ norm, the second one is
$O_{L^2(\Omega)}(\frac{1}{\tau})$. Applying Proposition \ref{vanka}
to the boundary value problem
$$
L(x,D)U_*=\widetilde fe^{\tau\Phi}\quad\mbox{in}\,\,\Omega, \quad
{\mathcal B}(x,D) U_*\vert_{\Gamma_0}=0,
$$
we construct a solution such that
$$
\Vert U_{*}e^{-\tau\varphi}\Vert_{W_2^{1,\tau}(\Omega)}\le C\Vert
f\Vert_{L^p(\Omega)}.
$$ The function $U^*-\tilde w$ solves the boundary value problem
(\ref{mimino}) and satisfies estimate $(\ref{mimino111}).$
 $\blacksquare$

Using Propositions \ref{Proposition 0} and \ref{Proposition 00},
we construct the last term $u_{-1}$ in complex geometric
optics solution which solves the equation
\begin{equation}\label{zad2!}
L_1(x,D) (e^{\tau\varphi}u_{-1})=-L_1(x,D) U,\quad\mathcal B(x,D)(e^{\tau\varphi}u_{-1})=- \mathcal B(x,D)U\quad\mbox{on}\,\,\Gamma_0
\end{equation} and  satisfies the estimate
\begin{equation}\label{mimino11}
\root\of{\vert\tau\vert} \Vert u_{-1} \Vert_{L^2(\Omega)} +
\frac{1}{\root\of{\vert\tau\vert} } \Vert \nabla u_{-1}
\Vert_{L^2(\Omega)}+\Vert u_{-1} \Vert_{W_2^{1,\tau}(\mathcal
O_{\epsilon})}=o(\frac{1}{\tau})\quad \mbox{as}\,\,\tau\rightarrow
+\infty.
\end{equation}
Finally we obtain a complex geometric optics solution in the form:
\begin{equation}\label{zad}
u_1(x)=U_{0,\tau}e^{\tau \Phi}+\widetilde U_{0,\tau} e^{\tau \overline
\Phi}-e^{\tau\Phi}\widetilde{\mathcal R}_{\tau, B_1}(q_1+\widetilde
q_1/\tau)-e^{\tau\overline\Phi}{\mathcal R}_{\tau, A_1}(q_2+\widetilde
q_2/\tau)+e^{\tau \varphi} u_{-1}.
\end{equation}

Obviously
\begin{equation}\label{zad22}
L_1(x,D)u_1=0\quad\mbox{in}\,\,\Omega,\quad {\mathcal B}(x,D) u_1\vert_{\Gamma_0}=0.
\end{equation}

Consider the operator
\begin{eqnarray}
{L}_2(x,{D})^{*} =4\partial_{ z} {\partial}_{\overline z}
-2{{A}^*_{2}}{\partial}_{\overline z}
-2{{B}^*_{2}}\partial_z+{Q^*_{2}} -2\partial_{\overline z}
{A}_{2}^*
-2\partial_z {B}^*_{2}\nonumber\\
=(2\partial_z-{{A}^*_{2}}) (2\partial_{\overline z}-{{B}^*_{2}})
+Q_1(2)
=(2\partial_{\overline z}-{{B}^*_{2}}) (2\partial_z-{{A}^*_{2}})
+Q_2(2).
                              \nonumber
\end{eqnarray}
Here
$$
Q_1(2)=Q_2^*-2\partial_{\bar z}A_2^*+A_2^*B_2^*,\quad Q_2(2)=Q_2^*-2\partial_zB_2^*+B_2^*A_2^*.
$$
Similarly we construct the complex geometric optics solutions to
the operator $L_2(x,D)^*.$
Let $(V_0,\widetilde V_0) \in C^{6+\alpha}(\overline \Omega)\times C^{6+\alpha}(\overline \Omega)$
be a solutions to the following boundary value problem:
\begin{equation}\label{ll1}
\mathcal M(x,D)(V_{0},\widetilde V_{0})=((2\partial_{ z}-{A_2^*})
V_{0},(2\partial_{\overline
z}-{B_2^*})\widetilde V_{0})=0\quad\mbox{in}\,\,\Omega, \quad  (V_{0}+\mbox{\bf I}\widetilde
V_{0})\vert_{\Gamma_0}=0,
\end{equation}
such that
\begin{equation}\label{iiii}
\lim_{x\rightarrow x_\pm}\frac{ \vert V_0(x)\vert}{\vert x-x_\pm\vert^{98}}
=\lim_{x\rightarrow x_\pm}\frac{ \vert \widetilde V_0(x)\vert}
{\vert x-x_\pm\vert^{98}}=0.
\end{equation}

Such a pair $(V_0,\widetilde V_0)$ exists due to Proposition
\ref{nikita}.
  Observe that
$$
{L}_2(x,{D})^{*}(\widetilde V_{0}e^{-\tau\Phi})=Q_1(2)\widetilde V_{0}e^{-\tau\Phi}\quad \mbox{in}\,\,\Omega,\quad
{L}_2(x,{D})^{*}(
V_{0}e^{-\tau\overline\Phi})=Q_2(2) V_{0} e^{-\tau\overline\Phi}\quad \mbox{in}\,\,\Omega.
$$

We set
\begin{equation}\label{loko1}
q_3=P_{- B^*_{2}}(Q_1(2)\widetilde V_{0})-M_{3},\quad q_4=T_{- A^*_{2}}(Q_2(2)V_{0})-M_{4},
\end{equation}
where $M_{3}\in Ker
(2\partial_{\overline z}-B_{2}^*)$ and
$M_{4}\in Ker (2\partial_{ z}-A_{2}^*)$ are chosen such that
\begin{equation}\label{lada1}
q_3(x)=q_4( x)=0, \quad
\forall x\in \mathcal H\quad\mbox{and}\,\, \lim_{x\rightarrow x_\pm}\frac{ \vert q_j(x)\vert}
{\vert x-x_\pm\vert^{98}}=0\quad j\in\{3,4\}.
\end{equation}

By (\ref{lada1}) the functions
$\frac{q_3}{2{\Phi '}},\frac{q_4}{2\partial_{\overline
z}\overline\Phi}$ belong to the space $C^2(\overline\Gamma_0).$
 Therefore  we  can
introduce the functions $V_{-1}, \widetilde V_{-1}$ as a solutions
to the following boundary value problem:
\begin{equation}
\mathcal M(x,D)(V_{-1},\widetilde V_{-1})=0\quad\mbox{in}\,\,\Omega,\quad
(V_{-1}+\mbox{\bf I}\widetilde
V_{-1})\vert_{\Gamma_0}=-(\mbox{\bf I}\frac{q_3}{2{\Phi '}}+\frac{q_4}
{2{\bar\Phi '}})-(0,(\vec e_2,\partial_\nu( V_0+\tilde V_0))/i\partial_\nu\psi),
\end{equation}

Let
$$
p_3=Q_1(2)(\frac{e_1q_{3}}{2{\Phi '}}
+\widetilde V_{-1})+L_2(x,D)^*(\frac{q_{3}e_2}{2{\Phi '}}),
p_4=Q_2(2)
(\frac{e_1q_{4}}{2\overline\Phi'}+ V_{-1})
+L_2(x,D)^*(\frac{q_{4}e_2}{2\overline\Phi'})
$$
and
$$
\widetilde q_4=(P_{-
B^*_{2}}p_4-\widetilde M_{3}), \quad
\widetilde q_3=(T_{-A^*_{2}} p_3-\widetilde M_{4}),
$$
where $\widetilde M_{3}\in Ker (2\partial_{\overline z}-B_{2}^*),
\widetilde M_{4}\in Ker
(2\partial_{ z}-A_{2}^*),$ and $(\widetilde q_{3},\widetilde
q_{4})$ are chosen such that
 \begin{equation}\label{lada}
\widetilde q_{3}( x)=\widetilde q_{4}( x)=0,\quad
 \forall x\in \mathcal H.
\end{equation}

The following asymptotic formula holds true:
\begin{proposition} \label{lp} Let $\overline G_\epsilon\cap \mbox{supp}\,e =\emptyset.$
There exist smooth functions $\widetilde m_{\pm,\tilde x}\in C^2(\overline
G_\epsilon), \tilde x\in \mathcal H$,
independent of $\tau$  such that
\begin{equation}
\widetilde{\mathcal R}_{-\tau,- A^*_{2}}(e_1(q_{3}+\widetilde
q_{3}/\tau))\vert_{\bar G_\epsilon}=\sum_{\tilde x\in\mathcal H}\frac{\widetilde m_{+,\tilde x}
e^{2i\tau(\psi-\psi(\widetilde
x))}}{\tau^2}+e^{2i\tau\psi}o_{W^1_2(\partial
\Omega)}(\frac{1}{\tau^2})\quad\mbox{as}\,\,\vert\tau\vert\rightarrow
+\infty
\end{equation}
and
\begin{equation}
{\mathcal R}_{-\tau,- B_{2}^*}(e_1(q_{4}+\widetilde
q_{4}/\tau))\vert_{\bar G_\epsilon}=\sum_{\tilde x\in\mathcal H}\frac{\widetilde m_{-,\tilde x}
e^{-2i\tau(\psi-\psi(\widetilde
x))}}{\tau^2}+e^{-2i\tau\psi}o_{W^1_2(\partial
\Omega)}(\frac{1}{\tau^2})\quad\mbox{as}\,\,\vert\tau\vert\rightarrow
+\infty.
\end{equation}
\end{proposition}

Using the functions $\tilde m_{\pm,\tilde x} $ we introduce functions
$\widetilde
 a_{\pm,\tilde x}, \widetilde  b_{\pm,\tilde x} \in C^2(\overline \Omega)$ which solve the boundary value problem

\begin{equation}
\mathcal M(x,D)(\widetilde a_{\pm,\tilde x},\widetilde
b_{\pm,\tilde x})=0\quad\mbox{in}\,\,\Omega,\quad (\widetilde a_{\pm,\tilde x}+\mbox{\bf I}\widetilde
b_{\pm,\tilde x})\vert_{\Gamma_0}=\pm(-\widetilde m_{\pm,\tilde x})\quad\forall\tilde x\in \mathcal H .
\end{equation}

By (\ref{lada}), there exists a pair $(V_{-2},\widetilde V_{-2})\in
W^1_2(\Omega)\times W^1_2(\Omega)$ which solves  the boundary value problem
\begin{equation}
\mathcal M(x,D)(V_{-2},\widetilde V_{-2})=0\quad
\mbox{in}\,\,\Omega,
\end{equation}
\begin{equation}
(V_{-2}+\mbox{\bf I}\widetilde V_{-2})\vert_{\Gamma_0}=-(\mbox{\bf I}\frac{\widetilde
q_3}{2{\Phi '}}+\frac{\widetilde q_4}{2
\overline\Phi'})-(0,\vec e_2,\partial_\nu(\tilde V_{-1}+\frac{e_2q_{3}}{2\Phi'}+V_{-1}+\frac{e_2q_{4}}{2\overline
\Phi'}))/i\partial_\nu\psi)
.
\end{equation}

We introduce  functions $V_{0,\tau},
\widetilde V_{0,\tau}$ by formulas
\begin{equation}\label{-1}
\tilde V_{0,\tau}=\tilde V_{0}+\frac{\tilde V_{-1}+\frac{e_2q_{3}}{2
\Phi'}}{\tau}+\frac{1}{\tau^2}(\sum_{\tilde x\in\mathcal H}( e^{2i\tau\psi(\widetilde x)}\widetilde
b_{+,\tilde x}+e^{-2i\tau\psi(\widetilde x)}\widetilde
b_{-,\tilde x})+\tilde V_{-2}+\frac{e_2\widetilde q_{3}}{2{\Phi '}})
\end{equation}
and
\begin{equation}\label{-2}
V_{0,\tau}=V_{0}+\frac{
V_{-1}+e_2q_{4}/2\overline
\Phi'}{\tau}+\frac{1}{\tau^2}(\sum_{\tilde x\in\mathcal H}( e^{2i\tau\psi(\widetilde x)}\widetilde
a_{+,\tilde x}+e^{-2i\tau\psi(\widetilde x)}\widetilde a_{-,\tilde x})+
V_{-2}+\frac{e_2\widetilde q_{4}}{2{\bar\Phi '}}).
\end{equation}
By (\ref{lada1}) and (\ref{lada}), the functions $V_{0,\tau}, \widetilde
V_{0,\tau}$ belong to $W^1_2(\Omega).$ After short computations, for any
$p\in (1,+\infty)$ we have
\begin{eqnarray}\label{90}
{L}_2(x,{D})^{*}\left(-e^{-\tau\Phi}\widetilde{\mathcal R}
_{-\tau,-A_{2}^*}(e_1(q_{3}+\frac{\widetilde
q_{3}}{\tau}))+\frac{e^{-\tau\Phi}e_2(q_{3}+\frac{\widetilde
q_{3}}{\tau})}{2\tau{\Phi '}}\right.\nonumber\\
\left.-e^{-\tau\overline\Phi}{\mathcal R}_{-\tau,- B_{2}^*}
(e_1(q_{4}+\frac{\widetilde
q_{4}}{\tau}))+\frac{e^{-\tau\overline\Phi}e_2(q_{4}+\frac{\widetilde
q_{4}}{\tau})}{2\tau{\bar\Phi '}}\right )
\nonumber\\
= -e^{-\tau\Phi}Q_2(2)\widetilde{\mathcal R}_{-\tau,-
A_{2}^*}(e_1(q_{3}+\frac{\widetilde
q_{3}}{\tau}))+e^{-\tau\Phi}L_2(x,D)^*(\frac{e_2(q_{3}+\frac{\widetilde
q_{3}}{\tau})}{2\tau{\Phi '}}) \nonumber \\
-e^{-\tau\overline\Phi}(Q_1(2){\mathcal
R}_{-\tau,-B_{2}^*}(e_1(q_{4}+\frac{\widetilde q_{4}}{\tau}))
+e^{-\tau\overline\Phi}L_2(x,D)^*(\frac{e_2(q_{4}+\frac{\widetilde
q_{4}}{\tau})}{2\tau{\bar\Phi '}})\nonumber\\
-e^{-\tau\Phi}Q_2(2)\frac{e_1q_{3}}{2\tau\Phi'}-e^{-\tau\overline\Phi} Q_1(2)\frac{e_1q_{4}}{2\tau\bar\Phi'}
-e^{-\tau\Phi}L_2(x,D)^*(\frac{e_2q_{3}}{2\tau{\Phi '}})\nonumber\\-e^{-\tau\overline\Phi}L_2(x,D)^*(\frac{e_2q_{4}}{2\tau{\bar\Phi '}})\nonumber
\\
-Q_1(2)(\tilde V_{-1}+\frac{\tilde V_{-2}}{\tau})e^{-\tau\Phi} -Q_2(2)(
V_{-1}+\frac{V_{-2}}{\tau})
e^{-\tau\overline\Phi}\nonumber\\
=-Q_1(2)(\tilde V_{-1}+\frac{\tilde V_{-2}}{\tau})e^{-\tau\Phi} -Q_2(2)(
V_{-1}+\frac{ V_{-2}}{\tau})
e^{-\tau\overline\Phi}+e^{-\tau\varphi}o_{L^p(\Omega)}(\frac{1}{\tau}).
\end{eqnarray}

Setting $V^*=V_{0,\tau}e^{-\tau \bar\Phi}+\widetilde V_{0,\tau}e^{-\tau
 \Phi}-e^{-\tau\Phi}\widetilde{\mathcal R}_{-\tau,-
A^*_{2}}(e_1(q_{3}+\frac{\widetilde
q_{3}}{\tau}))-e^{-\tau\overline\Phi}{\mathcal R}_{-\tau,
-B_{2}^*}(e_1(q_{4}+\frac{\widetilde q_{4}}{\tau}))$  for any
$p\in (1,\infty)$, we obtain that

\begin{equation}\label{nino}
L_2(x,D)^*V^*=e^{-\tau\varphi}o_{L^p(\Omega)}(\frac
1\tau)\quad\mbox{in}\,\,\Omega, \quad
{\mathcal B}(x,D) V^*\vert_{\Gamma_0}=e^{-\tau\varphi}\left(\begin{matrix} o_{W^1_2(\Gamma_0)}(\frac {1}{\tau^2})\\ O_{W^1_2(\Gamma_0)}(\frac {1}{\tau^2})\end{matrix}\right ).
\end{equation}
The first equality in (\ref{nino}) follows from (\ref{90}) and the second one can be obtained by argument similar to one used in  the proof of
Proposition \ref{Proposition 00}.

Using (\ref{nino}) and Proposition \ref{Proposition 0}, we construct the last term $v_{-1}$ in
complex geometric optics solution which solves
the boundary value problem
\begin{equation}\label{nino1}
L_2(x,D)^*v_{-1}=L_2(x,D)^*V^* \quad\mbox{in}\,\,\Omega, \quad
{\mathcal B}(x,D)v_{-1}\vert_{\Gamma_0}=-{\mathcal B}(x,D) V^* \end{equation}
and we obtain
\begin{equation}\label{Amimino11}
\root\of{\vert\tau\vert} \Vert v_{-1} \Vert_{L^2(\Omega)} +
\frac{1}{\root\of{\vert\tau\vert} } \Vert \nabla v_{-1}
\Vert_{L^2(\Omega)}+\Vert v_{-1} \Vert_{W_2^{1,\tau}(\mathcal
O_{\epsilon})}=o(\frac{1}{\tau})\quad \mbox{as}\,\, \tau\rightarrow+\infty.
\end{equation}
Finally we have a complex geometric optics solution for the
Schr\"odinger operator $L_2(x,D)^*$ in a form:
\begin{equation}\label{-3}
v=V_{1,\tau}e^{-\tau \bar\Phi}+\widetilde V_{1,\tau}e^{-\tau
\Phi}-e^{-\tau\Phi}\widetilde{\mathcal R}_{-\tau,-
A^*_{2}}(e_1(q_{3}+\frac{\widetilde
q_{3}}{\tau}))
-e^{-\tau\overline\Phi}{\mathcal R}_{-\tau,
-B_{2,}^*}(e_1(q_{4}+\frac{\widetilde q_{4}}{\tau}))
+v_{-1}e^{-\tau \varphi}.
\end{equation}

By (\ref{-3}), (\ref{nino}) and (\ref{nino1}), we have
\begin{equation}\label{-4}
L_2(x,D)^*v=0\quad\mbox{in}\,\,\Omega, \quad {\mathcal B}(x,D)  v\vert_{\Gamma_0}=0.
\end{equation}
\bigskip
\section{\bf Step 4: Asymptotic}
\bigskip
We introduce the following functionals
$$
\frak F_{\tau}u=\sum_{\tilde x\in \mathcal H}\frac{\pi}{2}\left(\frac{u(\widetilde x)}{\tau}-\frac{\partial_{zz}^2u(\widetilde x)}{2\Phi''(\tilde x)\tau^2}+\frac{\partial^2_{\overline z\overline z}u(\widetilde x)}{2\bar\Phi''(\tilde x)\tau^2}\right).
$$
 and
$$
\frak I_\tau r=\int_{\partial\Omega} r\frac{(\nu_1-i\nu_2)}
{2\tau {\Phi '}}e^{\tau(\Phi-\overline \Phi)}d\sigma
-\int_{\partial\Omega} \frac{(\nu_1-i\nu_2)}{{\Phi '}}\partial_z
\left (\frac{r}{2\tau^2 {\Phi '}}\right )e^{\tau(\Phi-\overline \Phi)} d\sigma.
$$

Using these  notations and the  fact that $\Phi$ is the harmonic function we rewrite the classical result  of theorem 7.7.5 of \cite{Her} as
\begin{proposition}\label{osel}
Let $\Phi(z)$ satisfies (\ref{zzz}), (\ref{mika}) and $u\in C^{5+\alpha}(\Omega),
\alpha\in (0,1)$ be some function.
Then the following asymptotic formula is true:
\begin{equation}\label{murzik9}
\int_{\Omega}ue^{\tau(\Phi-\overline \Phi)}dx=\frak F_{\tau}u+\frak J_{\tau}u
+ o\left(\frac{1}{\tau}\right)\quad \mbox{as}\,\,\tau\rightarrow +\infty.
\end{equation}
\end{proposition}

Denote
$$
\mbox{\bf H}(x,\partial_z,\partial_{\overline z})=2\mathcal A\partial_z+2\mathcal B\partial_{\bar z}+\mathcal Q,
$$ where $\mathcal A(x),\mathcal B(x)$ and $\mathcal Q(x)$ are some $2\times 2$ matrices .
We have
\begin{proposition}\label{lodka} Suppose that for any $U_{0,\tau}, \tilde U_{0,\tau}$  given by (\ref{zad1}), (\ref{zad2}) and $V_{0,\tau}, \tilde V_{0,\tau}$ given by (\ref{-1}), (\ref{-2}) with any function $\Phi$ which satisfies (\ref{zzz}), (\ref{mika})  we have
$$
\mathcal J_\tau=\int_\Omega (\mbox{\bf H}(x,\partial_z,\partial_{\overline z})U_{0,\tau},V_{0,\tau})dx=o(\frac 1\tau)\quad \mbox{as}\,\,\tau\rightarrow +\infty
$$
Then
\begin{eqnarray}\label{zaika}
J_\tau=\sum_{k=-1}^1\tau^k J_k+\frac 1\tau((J_++\sum_{\tilde x\in \mathcal H}I_{+,\Phi}(\tilde x)+K_+)e^{2\tau i\psi(\tilde x)}+(J_-+ \sum_{\tilde x\in \mathcal H}I_{-,\Phi}(\tilde x)+K_-)e^{-2\tau i\psi(\tilde x)})\nonumber\\+\int_{\tilde\Gamma}((\nu_1-i\nu_2)(\mathcal A U_0e^{\tau\Phi},V_0e^{-\tau\bar\Phi})+(\nu_1+i\nu_2)(\mathcal B \tilde U_0e^{\tau\bar\Phi},\tilde V_0e^{-\tau\Phi}))d\sigma\nonumber\\
+\frak J_\tau(q_1 , T^*_{B_1}B_1^*\mathcal A^* V_{0,\tau}-\mathcal A^*V_{0,\tau}+2T_{B_1}^*\partial_z \mathcal B^* V_{0,\tau}+T_{B_1}^*(\mathcal B^*(A^*_2V_{0,\tau}-2\tau\bar\Phi'V_{0,\tau}))\nonumber\\
+\frak J_{-\tau} (q_2,P_{A_1}^*(2\mathcal A^*\tilde V_{0,\tau} -\tau\Phi'2\mathcal A^*\tilde V_{0,\tau})-\mathcal B^*\tilde V_{0,\tau}+P^*_{A_1}(A_1^*\mathcal B^*\tilde V_{0,\tau}))\nonumber\\
-\frak J_{-\tau}( q_3,T_{-B_2^*}^*(2\mathcal A\partial_z \tilde U_{0,\tau}+2\mathcal B(\partial_z\tilde U_{0,\tau}+\tau\bar\Phi'\tilde U_{0,\tau})))\nonumber\\
-2\frak J_{\tau}(P_{-A_2^*}^*(\mathcal A (\partial_zU_{0,\tau}+\tau\Phi' U_{0,\tau}))+\mathcal B \partial_{\bar z} U_{0,\tau}, q_4)+o(\frac 1\tau)\quad\mbox{as}\,\,\tau\rightarrow +\infty,
\end{eqnarray}
where
\begin{eqnarray}
J_+=\frac{\pi}{2}(-(2\partial_z\mathcal AU_0, V_0)-(\mathcal AU_0, B_2^*V_0) -(2\mathcal B A_1 U_0, V_0)
-\frac 12(Q_1(1)U_0, T^*_{B_1}(\mathcal B^*V_0))\nonumber\\-(Q_1(2)\tilde U_0,P^*_{-A^*_2}(\mathcal AU_0)))+(\mathcal Q U_0,V_0)),
\end{eqnarray}
\begin{eqnarray}
J_-=\frac{\pi}{2}((2\mathcal A\partial_z \tilde U_0, \tilde V_0)-(2\partial_{\bar z}\mathcal B\tilde U_0, \tilde V_0) -(2\mathcal B\tilde  U_0, \partial_{\bar z}\tilde V_0)\nonumber\\
-((\partial_{ z} \tilde q_1, P^*_{A^*_1}(\mathcal A^* \tilde V_0))+(\partial_{\bar z}\tilde  q_2, T^*_{-B^*_2}(\mathcal B\tilde U_0)))+ (\mathcal Q\tilde U_0,\tilde V_0)),
\end{eqnarray}
\begin{eqnarray}\label{lin1}
I_{+,\Phi}(x)=\sum_{\tilde x\in\mathcal H}\int_{\partial\Omega}\left\{ (\nu_1-i\nu_2)((2\mathcal B b_{+,\tilde x}\bar\Phi', V_0)+
(2\bar\Phi' U_0, \tilde a_{+,\tilde x}))\right.\nonumber\\\left.+(\nu_1+i\nu_2)
((2 a_{+,\tilde x}\Phi', \tilde V_0)
+(2\Phi'\tilde U_0, \tilde b_{+,\tilde x}))\right\}d\sigma,
\end{eqnarray}

\begin{eqnarray}\label{lin2}
I_{-,\Phi}(x)=\sum_{\tilde x\in\mathcal H}\int_{\partial\Omega}\left \{(\nu_1-i\nu_2)((2\mathcal B b_{+,\tilde x}\bar\Phi', V_0)+
(2\bar\Phi' U_0, \tilde a_{+,\tilde x}))\right.\nonumber\\\left.+(\nu_1+i\nu_2)
((2 a_{+,\tilde x}\Phi', \tilde V_0)
+(2\Phi'\tilde U_0, \tilde b_{+,\tilde x}))\right\}d\sigma,
\end{eqnarray}
\begin{eqnarray}\label{zubilo}
K_+=
\tau\frak F_\tau(q_1 , T^*_{B_1}B_1^*\mathcal A^* V_0-\mathcal A^*V_0+2T_{B_1}^*\partial_z \mathcal B^* V_0+T_{B_1}^*(\mathcal B^*(A^*_2V_0-2\tau\bar\Phi'V_0)))
\nonumber\\
-2\tau\frak F_{\tau}(P_{-A_2^*}^*(\mathcal A (\partial_zU_0+\tau\Phi' U_0)+\mathcal B \partial_{\bar z} U_{0,\tau}), q_4).
\end{eqnarray}
\end{proposition}

{\bf Proof.}
Denote
\begin{equation}\label{pravda3}
U_1=-\widetilde{\mathcal R}_{\tau, B_1}(e_1(q_1+\widetilde
q_1/\tau)),\quad \tilde U_1= -{\mathcal R}_{\tau,
A_1}(e_1(q_2+\widetilde q_2/\tau)),
\end{equation}
\begin{equation}\label{pravda4}\tilde V_1= -\widetilde{\mathcal R}_{-\tau,-
A^*_{2}}(e_1(q_{3}+\frac{\widetilde
q_{3}}{\tau})),
\quad
 V_1=
-{\mathcal R}_{-\tau,
-B_{2}^*}(e_1(q_{4}+\frac{\widetilde q_{4}}{\tau})).
\end{equation}
Integrating by parts and using Proposition \ref{osel}, we obtain

\begin{eqnarray}
\mathcal M_1=\int_\Omega (2\mathcal A\partial_z (U_{0,\tau}e^{\tau\Phi}) +2\mathcal B\partial_{\bar z} (U_{0,\tau}e^{\tau\Phi}), V_{0,\tau}e^{-\tau\bar\Phi})dx=\nonumber\\
\int_\Omega ((-2\partial_z\mathcal AU_{0,\tau}e^{\tau\Phi}, V_{0,\tau}e^{-\tau\bar\Phi})-(2\mathcal A U_{0,\tau}e^{\tau\Phi}, \partial_zV_{0,\tau}e^{-\tau\bar\Phi}) +(2\mathcal B\partial_{\bar z} U_{0,\tau}e^{\tau\Phi}, V_{0,\tau}e^{-\tau\bar\Phi}))dx\nonumber\\
+\int_{\partial\Omega}(\nu_1-i\nu_2)(\mathcal A U_{0,\tau}e^{\tau\Phi},V_{0,\tau}e^{-\tau\bar\Phi})d\sigma\nonumber=\\
\frak F_\tau(-(2\partial_z\mathcal AU_0, V_0)-(2\mathcal AU_0, \partial_zV_0) +(2\mathcal B\partial_{\bar z} U_0, V_0))\nonumber\\+
\frak I_\tau(-(2\partial_z\mathcal AU_{0,\tau}, V_{0,\tau})-(2\mathcal AU_{0,\tau}, \partial_zV_{0,\tau}) +(2\mathcal B\partial_{\bar z} U_{0,\tau}, V_{0,\tau}))\nonumber\\
+\int_{\tilde \Gamma}(\nu_1-i\nu_2)(\mathcal A U_0,V_0)e^{\tau(\Phi-\bar\Phi)}d\sigma+\kappa_{0,0}+\frac{\kappa_{0,-1}}{\tau}+o(\frac 1\tau),
\end{eqnarray}
where $\kappa_{0,j}$ are  some constants independent of $\tau.$

Integrating by parts we obtain that there exist constants  $\kappa_{1,j}$  independent of $\tau$ such that
\begin{eqnarray}\label{sor1}
\int_\Omega (2\mathcal A\partial_z (\tilde U_{0,\tau}e^{\tau\bar\Phi}) +2\mathcal B\partial_{\bar z} (\tilde U_{0,\tau}e^{\tau\bar \Phi}), V_{0,\tau}e^{-\tau\bar\Phi})dx=\nonumber\\
(2\partial_z\mathcal A\tilde U_{0,\tau},V_{0,\tau})_{L^2(\Omega)}+(2\mathcal B(\partial_z\tilde U_{0,\tau}+\tau\bar \Phi'\tilde U_{0,\tau}),V_{0,\tau})_{L^2(\Omega)}=\nonumber\\
\tau\kappa_{1,1}+\kappa_{1,0}+\frac{\kappa_{1,-1}}{\tau}+\frac 1\tau\sum_{\tilde x\in\mathcal H}(e^{2i\tau\psi(\tilde x)}(2\mathcal B b_{+,\tilde x}\bar\Phi', V_0)_{L^2(\Omega)}+e^{-2i\tau\psi(\tilde x)}(2\mathcal B b_{-,\tilde x}\bar\Phi', V_0)_{L^2(\Omega)})\nonumber\\
+\frac 1\tau\sum_{\tilde x\in\mathcal H}(e^{2i\tau\psi(\tilde x)}(2\mathcal B\bar\Phi' U_0, \tilde a_{+,\tilde x})_{L^2(\Omega)}+e^{-2i\tau\psi(\tilde x)}(2\mathcal B\bar\Phi'U_0, \tilde a_{-,\tilde x})_{L^2(\Omega)})+o(\frac 1\tau).
\end{eqnarray}
Since for any $\tilde x$ from $\mathcal H$
$$(2\mathcal B\bar\Phi' U_0, \tilde a_{\pm,\tilde x})=4\partial_z(\bar\Phi' U_0, \tilde a_{\pm,\tilde x}),\quad\mbox{and}\,\,(2\mathcal B b_{\pm,\tilde x}\bar\Phi', V_0)=4\partial_z(b_{\pm,\tilde x}\bar\Phi', V_0)\quad\mbox{in}\,\,\Omega
$$ from (\ref{sor1}) we have
\begin{eqnarray}\label{sor2}
\mathcal M_2=\int_\Omega (2\mathcal A\partial_z (\tilde U_{0,\tau}e^{\tau\bar\Phi}) +2\mathcal B\partial_{\bar z} (\tilde U_{0,\tau}e^{\tau\bar \Phi}), V_{0,\tau}e^{-\tau\bar\Phi})dx=\nonumber\\
\tau\kappa_{1,1}+\kappa_{1,0}+\frac{\kappa_{1,-1}}{\tau}+\int_{\partial\Omega}\sum_{\tilde x\in\mathcal H}\frac {(\nu_1-i\nu_2)}{\tau}(e^{2i\tau\psi(\tilde x)}(2\mathcal B b_{+,\tilde x}\bar\Phi', V_0)+e^{-2i\tau\psi(\tilde x)}(2\mathcal B b_{-,\tilde x}\bar\Phi', V_0))d\sigma\nonumber\\
+\int_{\partial\Omega}\sum_{\tilde x\in\mathcal H}\frac {(\nu_1-i\nu_2)}{\tau}(e^{2i\tau\psi(\tilde x)}(2\bar\Phi' U_0, \tilde a_{+,\tilde x})+e^{-2i\tau\psi(\tilde x)}(2\bar\Phi'U_0, \tilde a_{-,\tilde x}))d\sigma+o(\frac 1\tau).
\end{eqnarray}
Integrating by parts we obtain that there exist  constants $\kappa_{2,j}$  independent of $\tau$ such that
\begin{eqnarray}\label{sor3}
\int_\Omega (2\mathcal A\partial_z (U_{0,\tau}e^{\tau\Phi}) +2\mathcal B\partial_{\bar z} (U_{0,\tau}e^{\tau\Phi}),\tilde V_{0,\tau}e^{-\tau\Phi})dx=\nonumber\\
(2\mathcal A(\partial_z U_{0,\tau}+\tau\Phi' U_{0,\tau}) +2\mathcal B\partial_{\bar z} U_{0,\tau},\tilde V_{0,\tau})_{L^2(\Omega)}=\nonumber\\
\tau\kappa_{2,1}+\kappa_{1,0}+\frac{\kappa_{2,-1}}{\tau}+
\frac 1\tau\sum_{\tilde x\in\mathcal H}(e^{2i\tau\psi(\tilde x)}(2\mathcal A a_{+,\tilde x}\Phi', \tilde V_0)_{L^2(\Omega)}+e^{-2i\tau\psi(\tilde x)}(2\mathcal A a_{-,\tilde x}\Phi',\tilde  V_0)_{L^2(\Omega)})\nonumber\\
+\frac 1\tau\sum_{\tilde x\in\mathcal H}(e^{2i\tau\psi(\tilde x)}(2\mathcal A \Phi'\tilde U_0, \tilde b_{+,\tilde x})_{L^2(\Omega)}+e^{-2i\tau\psi(\tilde x)}(2\mathcal A \Phi'\tilde U_0,\tilde  b_{-,\tilde x})_{L^2(\Omega)})+o(\frac 1\tau).
\end{eqnarray}
Since  for any $\tilde x$ from $\mathcal H$
$$
(2\mathcal A a_{\pm,\tilde x}\Phi', \tilde V_0)=4\partial_{\bar z}( a_{\pm,\tilde x}\Phi', \tilde V_0)\quad\mbox{and}\quad (2\mathcal A \Phi'\tilde U_0, \tilde b_{\pm,\tilde x})=4\partial_{\bar z}( \Phi'\tilde U_0, \tilde b_{\pm,\tilde x})\quad\mbox{in}\,\,\Omega
$$ we obtain from (\ref{sor3})

\begin{eqnarray}\label{sor33}
\mathcal M_3=\int_\Omega (2\mathcal A\partial_z (U_{0,\tau}e^{\tau\Phi}) +2\mathcal B\partial_{\bar z} (U_{0,\tau}e^{\tau\Phi}),\tilde V_{0,\tau}e^{-\tau\Phi})dx=\nonumber\\
\tau\kappa_{2,1}+\kappa_{1,0}+\frac{\kappa_{2,-1}}{\tau}+\int_{\partial\Omega}(\nu_1+i\nu_2)
\frac 1\tau\sum_{\tilde x\in\mathcal H}(e^{2i\tau\psi(\tilde x)}(2 a_{+,\tilde x}\Phi', \tilde V_0)+e^{-2i\tau\psi(\tilde x)}(2 a_{-,\tilde x}\Phi',\tilde  V_0))d\sigma\nonumber\\
+\int_{\partial\Omega}(\nu_1+i\nu_2)\frac 1\tau\sum_{\tilde x\in\mathcal H}(e^{2i\tau\psi(\tilde x)}(2\Phi'\tilde U_0, \tilde b_{+,\tilde x})+e^{-2i\tau\psi(\tilde x)}(2\Phi'\tilde U_0,\tilde  b_{-,\tilde x})) d\sigma+o(\frac 1\tau).
\end{eqnarray}
Integrating by parts, using  (\ref{-55!}) and Proposition \ref{osel},
we obtain  that there exists some constants $\kappa_{3,j}$ independent
of $\tau$ such that

\begin{eqnarray}
\mathcal M_4=\int_\Omega (2\mathcal A\partial_z (\tilde U_{0,\tau}e^{\tau\bar\Phi}) +2\mathcal B\partial_{\bar z} (\tilde U_{0,\tau}e^{\tau\bar\Phi}), \tilde V_{0,\tau}e^{-\tau\Phi})dx=\nonumber\\
\int_\Omega ((2\mathcal A\partial_z \tilde U_{0,\tau}e^{\tau\bar\Phi}, \tilde V_{0,\tau}e^{\tau\Phi})-(2\partial_{\bar z}\mathcal B\tilde U_{0,\tau}e^{\tau\bar\Phi},\tilde V_{0,\tau}e^{-\tau\Phi}) -(2\mathcal B\tilde U_{0,\tau}e^{\tau\Phi}, \partial_{\bar z}\tilde V_{0,\tau}e^{\tau\bar\Phi}))dx\nonumber\\
+\int_{\partial\Omega}(\nu_1+i\nu_2)(\mathcal B \tilde U_{0,\tau}e^{\tau\bar\Phi},\tilde V_{0,\tau}e^{-\tau\Phi})d\sigma\nonumber=\\
\frak F_{-\tau}( (2\mathcal A\partial_z \tilde U_0, \tilde V_0)-(2\partial_{\bar z}\mathcal B\tilde U_0, \tilde V_0) -(2\mathcal B\tilde  U_0, \partial_{\bar z}\tilde V_0))\nonumber\\+
\frak I_{-\tau}(( (2\mathcal A\partial_z \tilde U_{0,\tau}, \tilde V_{0,\tau})-(2\partial_{\bar z}\mathcal B\tilde U_{0,\tau},\tilde  V_{0,\tau}) -(2\mathcal B \tilde U_{0,\tau}, \partial_{\bar z}\tilde V_{0,\tau}))\nonumber\\
+\int_{\tilde\Gamma}(\nu_1+i\nu_2)(\mathcal B \tilde U_0e^{\tau\bar\Phi},\tilde V_0e^{-\tau\Phi})d\sigma+\kappa_{3,1}+\frac{\kappa_{3,-1}}{\tau}+o(\frac 1\tau).
\end{eqnarray}

Integrating by parts and using Proposition \ref{osel} we obtain

\begin{eqnarray}
\mathcal M_5=\int_\Omega (2\mathcal A\partial_z (U_1e^{\tau\Phi}) +2\mathcal B\partial_{\bar z} (U_1e^{\tau\Phi}), V_{0,\tau}e^{-\tau\bar\Phi})dx=\\
\int_\Omega (\mathcal A (-B_1U_1-q_1)e^{\tau\Phi} -2\partial_{\bar z}\mathcal B (U_1e^{\tau\Phi}), V_{0,\tau}e^{-\tau\bar\Phi})dx+\nonumber\\
\int_{\partial\Omega}(\nu_1+i\nu_2)(\mathcal BU_1,V_0)e^{\tau(\Phi-\bar\Phi)}d\sigma-(2\mathcal B U_1,\partial_{\bar z}(V_{0,\tau}e^{\tau(\Phi-\bar\Phi)}))_{L^2(\Omega)}=\nonumber\\
\int_\Omega (\mathcal A (B_1T_{B_1}(e^{\tau(\Phi-\bar\Phi)}q_1)-q_1)e^{\tau(\Phi-\bar\Phi)}, V_{0,\tau}) +2\partial_z\mathcal B (T_{B_1}(e^{\tau(\Phi-\bar\Phi)}q_1)), V_{0,\tau})dx+\nonumber\\
(\mathcal B T_{B_1}(e^{\tau(\Phi-\bar\Phi)}q_1), A^*_2 V_{0,\tau}-2\tau\bar\Phi' V_{0,\tau})_{L^2(\Omega)}
+\int_{\partial\Omega}(\nu_1+i\nu_2)(\mathcal BU_1,V_{0,\tau})e^{\tau(\Phi-\bar\Phi)}d\sigma
=\nonumber\\
\frak F_\tau(q_1 , T^*_{B_1}B_1^*\mathcal A^* V_0-\mathcal A^*V_0+2T_{B_1}^*(\partial_z \mathcal B^* V_0)+T_{B_1}^*(\mathcal B^*(A^*_2V_0-2\tau\bar\Phi'V_0)))\nonumber\\
+\frak I_\tau(q_1 , T^*_{B_1}B_1^*\mathcal A^* V_{0,\tau}-\mathcal A^*V_{0,\tau}+2T_{B_1}^*(\partial_z \mathcal B^* V_{0,\tau})+T_{B_1}^*(\mathcal B^*(A^*_2V_{0,\tau}-2\tau\bar\Phi'V_{0,\tau}))\nonumber\\
+\int_{\partial\Omega}(\nu_1+i\nu_2)(\mathcal BU_1,V_{0,\tau})e^{\tau(\Phi-\bar\Phi)}d\sigma+o(\frac 1\tau).\nonumber
\end{eqnarray}

After integration by parts we have
\begin{eqnarray}
\int_\Omega (2\mathcal A\partial_z (U_1e^{\tau\Phi}) +2\mathcal B\partial_{\bar z} (U_1e^{\tau\Phi}), \tilde V_{0,\tau}e^{-\tau\Phi})dx=\nonumber\\
\int_\Omega (\mathcal A (-B_1U_1-q_1) -2\partial_{\bar z}\mathcal B U_1, \tilde V_{0,\tau})dx+\nonumber\\
(2\mathcal B U_1,\partial_{\bar z}\tilde V_0)_{L^2(\Omega)}+\int_{\partial\Omega}(\nu_1+i\nu_2)(\mathcal B U_1,\tilde V_{0,\tau})d\sigma .\nonumber
\end{eqnarray}
Using (\ref{pravda3}) and Proposition 8 of \cite{IY} we obtain that
\begin{equation}
\mathcal M_6=\int_\Omega (2\mathcal A\partial_z (U_1e^{\tau\Phi}) +2\mathcal B\partial_{\bar z} (U_1e^{\tau\Phi}), \tilde V_{0,\tau}e^{-\tau\Phi})dx=-\int_\Omega (\mathcal A q_1,\tilde V_{0,\tau})dx+o(\frac 1{\tau^2})\quad\mbox{as}\quad\tau\rightarrow +\infty.
\end{equation}
Integrating by parts and using Proposition \ref{osel} we have

\begin{eqnarray}\mathcal M_7=
\int_\Omega (2\mathcal A\partial_z (U_{0,\tau}e^{\tau\Phi}) +2\mathcal B\partial_{\bar z} (U_{0,\tau}e^{\tau\Phi}), V_1e^{-\tau\bar\Phi})dx=\\
2\int_\Omega (\mathcal A (\partial_zU_{0,\tau}+\tau\Phi' U_{0,\tau})e^{\tau\Phi} +\mathcal B \partial_{\bar z} U_{0,\tau}e^{\tau\Phi}, V_1e^{-\tau\bar\Phi})dx=\nonumber\\
-2\int_\Omega (P_{-A_2^*}^*(\mathcal A (\partial_zU_0+\tau\Phi' U_0)+\mathcal B \partial_{\bar z} U_{0,\tau}), q_4e^{\tau(\Phi-\bar\Phi)})dx=\nonumber\\
 -2\frak F_{\tau}(P_{-A_2^*}^*(\mathcal A (\partial_zU_0+\tau\Phi' U_0)+\mathcal B \partial_{\bar z} U_{0}), q_4)\nonumber+\\
  -2\frak I_{\tau}(P_{-A_2^*}^*(\mathcal A (\partial_zU_{0,\tau}+\tau\Phi' U_{0,\tau})+\mathcal B \partial_{\bar z} U_{0,\tau}), q_4)+o(\frac 1\tau)\quad\mbox{as}\quad\tau\rightarrow +\infty.\nonumber
\end{eqnarray}

Integrating by parts and using Proposition  8 of \cite{IY} we have
\begin{eqnarray}
\mathcal M_8=\int_\Omega (2\mathcal A\partial_z (U_{0,\tau}e^{\tau\Phi}) +2\mathcal B\partial_{\bar z} (U_{0,\tau}e^{\tau\Phi}), \tilde V_1e^{-\tau\Phi})dx=\nonumber\\
\int_\Omega( (-2\partial_z\mathcal A U_0 +\mathcal B \partial_{\bar z} U_0, \tilde V_1)-(\mathcal A U_0,-B_2^*\tilde V_1-q_3))dx\nonumber\\
+\int_{\partial\Omega}(\nu_1-i\nu_2)(\mathcal AU_0,\tilde V_1)d\sigma=-\int_\Omega(\mathcal A U_{0,\tau}, q_3)dx+o(\frac{1}{\tau})\quad\mbox{as}\quad\tau\rightarrow +\infty
\end{eqnarray}

and

\begin{eqnarray}
\mathcal M_9=\int_\Omega (2\mathcal A\partial_z (\tilde U_1e^{\tau\bar\Phi}) +2\mathcal B\partial_{\bar z} (\tilde U_1e^{\tau\bar\Phi}),  V_{0,\tau}e^{-\tau\bar\Phi})dx=\nonumber\\
\int_\Omega [(\tilde U_1, -\partial_z(2\mathcal A^* V_{0,\tau}))+(\mathcal B(-A_1\tilde U_1-q_2), V_{0,\tau})]dx\nonumber\\
+\int_{\partial\Omega}(\nu_1-i\nu_2)(\mathcal A\tilde U_1, V_0)d\sigma=-\int_\Omega (\mathcal B q_2,V_{0,\tau})dx+o(\frac{1}{\tau})\quad\mbox{as}\quad\tau\rightarrow +\infty.
\end{eqnarray}

Integrating by parts and using Proposition \ref{osel} we obtain

\begin{eqnarray}
\mathcal M_{10}=\int_\Omega (2\mathcal A\partial_z (\tilde U_1e^{\tau\bar\Phi}) +2\mathcal B\partial_{\bar z} (\tilde U_1e^{\tau\bar\Phi}), \tilde  V_{0,\tau}e^{-\tau\Phi})dx=\\
\int_\Omega ((\tilde U_1, -\partial_z(2\mathcal A^* \tilde V_{0,\tau})+\tau\Phi'2\mathcal A^*\tilde V_{0,\tau})+(\mathcal B(-A_1\tilde U_1- q_2),\tilde V_{0,\tau})e^{\tau(\bar\Phi-\Phi)})dx+\nonumber\\
+\int_{\partial\Omega}(\nu_1-i\nu_2)(\mathcal A\tilde U_1,\tilde V_{0,\tau})e^{\tau(\bar\Phi-\Phi)}d\sigma=\nonumber\\
\int_\Omega( q_2,P_{A_1}^*(2\partial_z(\mathcal A^*\tilde V_{0,\tau}) -2\tau\Phi'\mathcal A^*\tilde V_0)-\mathcal B^*\tilde V_0-P^*_{A_1}(A_1^*\mathcal B^*\tilde V_0)))e^{\tau(\bar\Phi-\Phi)}dx\nonumber\\
+\int_{\partial\Omega}(\nu_1-i\nu_2)(\mathcal A\tilde U_1,\tilde V_0)e^{\tau(\bar\Phi-\Phi)}d\sigma=\nonumber\\
\frak F_{-\tau} ( q_2,P_{A_1}^*(2\partial_z(\mathcal A^*\tilde V_0) -\tau\Phi'2\mathcal A^*\tilde V_0)-\mathcal B^*\tilde V_0+P^*_{A_1}(A_1^*\mathcal B^*\tilde V_0))\nonumber\\
+\frak I_{-\tau} (q_2,P_{A_1}^*(2\partial_z(\mathcal A^*\tilde V_{0,\tau}) -\tau\Phi'2\mathcal A^*\tilde V_{0,\tau})-\mathcal B^*\tilde V_{0,\tau}+P^*_{A_1}(A_1^*\mathcal B^*\tilde V_{0,\tau}))+o(\frac{1}{\tau})\quad\mbox{as}\quad\tau\rightarrow +\infty.\nonumber
\end{eqnarray}

By (\ref{soika}) and Proposition \ref{osel} we obtain
\begin{eqnarray}
\mathcal M_{11}=\int_\Omega (2\mathcal A\partial_z (\tilde U_{0,\tau}e^{\tau\bar\Phi}) +2\mathcal B\partial_{\bar z} (\tilde U_{0,\tau}e^{\tau\bar\Phi}), \tilde  V_1e^{-\tau\Phi})dx=\\
\int_\Omega ((2\mathcal A\partial_z \tilde U_{0,\tau}+2\mathcal B(\partial_{\bar z}\tilde U_{0,\tau}+\tau\bar\Phi'\tilde U_{0,\tau}),\tilde V_1)e^{\tau(\bar\Phi-\Phi)}dx=\nonumber\\
-\int_\Omega( q_3,T_{-B_2^*}^*(2\mathcal A\partial_z \tilde U_{0,\tau}e^{\tau\bar\Phi}+2\mathcal B(\partial_z\tilde U_{0,\tau}+\tau\bar\Phi'\tilde U_{0,\tau}))e^{\tau(\bar\Phi-\Phi)}dx=\nonumber\\
-\frak F_{-\tau}( q_3,T_{-B_2^*}^*(2\mathcal A\partial_z \tilde U_{0}+2\mathcal B(\partial_z\tilde U_{0}+\tau\bar\Phi'\tilde U_{0})))\nonumber\\
-\frak J_{-\tau}( q_3,T_{-B_2^*}^*(2\mathcal A\partial_z \tilde U_{0,\tau}+2\mathcal B(\partial_z\tilde U_{0,\tau}+\tau\bar\Phi'\tilde U_{0,\tau})))+o(\frac 1\tau)\quad\mbox{as}\quad\tau\rightarrow +\infty.\nonumber
\end{eqnarray}

By Proposition  \ref{osel} we there exist constants $\kappa_{4,j}$ independent of $\tau$ such that

\begin{eqnarray}
\mathcal M_{12}=\int_\Omega ( U_{0,\tau}e^{\tau\bar \Phi}+ \tilde  U_{0,\tau}e^{\tau\Phi},  V_{0,\tau}e^{-\tau\bar \Phi}+ \tilde  V_{0,\tau}e^{-\tau\Phi})dx=\\ \kappa_{4,0}+\kappa_{4,-1}/\tau+
\frac{ \pi}{ 2\tau}((QU_0,V_0)(\tilde x)e^{2i\tau\psi(\tilde x)}+(Q\tilde U_0,\tilde V_0)(\tilde x)e^{-2i\tau\psi(\tilde x)})+o(\frac{1}{\tau})\quad\mbox{as}\quad\tau\rightarrow +\infty.\nonumber
\end{eqnarray}
Since $\mathcal J_\tau=\sum_{k=1}^{12}\mathcal M_k$ the proof of
the proposition is complete.
$\blacksquare$

From Proposition \ref{lodka} we  obtain:

\begin{proposition}\label{knopka} Under assumptions  of
Proposition \ref{lodka} the following equality holds true
\begin{equation}\label{ppp}
T^*_{B_1}(\mathcal B^*\bar\Phi'V_0)=P^*_{-A^*_2}(\mathcal A\Phi'V_0) =P^*_{A_1}(\mathcal B\Phi'\tilde V_0)=T^*_{-B^*_2}(\mathcal A^*\bar \Phi'\tilde U_0)\quad \mbox{on }\,\,\tilde\Gamma
\end{equation}
and
\begin{equation}\label{pppp}
T^*_{B_1}(\mathcal B^*V_0)=P^*_{-A^*_2}(\mathcal AV_0) =P^*_{A_1}(\mathcal B\tilde V_0)=T^*_{-B^*_2}(\mathcal A^*\tilde U_0)\quad \mbox{on }\,\,\tilde\Gamma.
\end{equation}
\end{proposition}

{\bf Proof.}
we construct the  function $r$ with domain $\partial\Omega$  i in the following way.
Let us parameterize the boundary of simply connected  domain $\Omega$ clockwise. We remind that without the loss of generality  we one can assume that $\tilde \Gamma$ is an ark with the endpoints $ x_-$ and $ x_+.$  Let $x^*$ be an arbitrary point from $\tilde \Gamma$. On $\Gamma_0$ we set the function $r$ be equal  zero. From $x_-$ till $x^*$ this is the  strictly increasing function and  form $x^*$ to $ x_+$  the function $r$  is strictly decreasing. At $x^*$ the function $r$ has nondegenerate critical point and there are no critical points on $( x_-,x_+)$ except of $x^*.$  Finally, at point $ x_-$ we assume that the right tangential derivative of order five of the function $r$ is not equal to zero and  at point $ x_+$ we assume that the left tangential derivative of order five of the function $r$ is not equal to zero. Let $\tilde\psi$  be a  harmonic function in $\bar \Omega$ such that $\psi\vert_{\partial\Omega}=r.$ Since domain $\Omega$ is simply connected  there exists a smooth harmonic function $\tilde \varphi$ such that the function $\tilde \Phi=\tilde \varphi+i\tilde\psi$ is holomorphic in the domain $\Omega$.
Moreover, after possible small perturbation of the function $r$ with support near point where  $r$ is strictly increasing we may assume that
$$
\partial_{\nu}\tilde\varphi( x_\pm) \ne 0 \quad \mbox{and}\quad \partial_{\nu}\tilde\varphi(x^*) \ne 0.
$$
Then the function $\tilde \Phi$ does not have a critical points on $\bar\Omega.$ Consider the complex geometric optics solution constructed with the phase function $\tilde \Phi=\tilde\varphi+i\tilde\psi$ instead of $\Phi$ and  the pair $(\frac{\Phi'}{\tilde \Phi'}U_0, \frac{\bar\Phi'}{\overline{\tilde \Phi'}}\tilde U_0)$ instead of $(U_0,\tilde U_0).$
By Proposition \ref{lodka}, the equality (\ref{zaika}) holds true. Since there are no critical points of the function $\tilde\Phi$ on $\bar\Omega$  the terms $I_\pm , K_\pm, J_\pm$ are equal to zero.  Let $e\in C^5(\partial\Omega)$ be a a function with support concentrated near points $ x_\pm$ and equal to one in some neighborhood of points $x_\pm.$  By (\ref{xoxo1u}),  (\ref{iiii})  and
the fact that the left or the right tangential derivative of order five of the function $\psi$ at points $x_\pm$ is not equal zero using the stationary phase argument we have
\begin{equation}\label{gop1}
\int_{\partial\Omega}e((\nu_1-i\nu_2)\mathcal AU_0,V_0)e^{\tau(\Phi-\bar\Phi)}d\sigma=c_1+o(\frac 1\tau)\quad \mbox{as}\,\,\tau\rightarrow +\infty.
\end{equation}
and
\begin{equation}\label{gop2}
\int_{\partial\Omega}e((\nu_1+i\nu_2)\mathcal BU_0,V_0)e^{\tau(\Phi-\bar\Phi)}d\sigma=c_2+o(\frac 1\tau)\quad \mbox{as}\,\,\tau\rightarrow +\infty.
\end{equation}
By (\ref{bin1}), (\ref{lada1}) and the fact that left or right tangential derivative of order five of the function $\psi$ at points $x_\pm$ is not equal zero using  using theorem 7.7.1 of \cite{Her} we obtain
\begin{equation}\label{gop3}
\int_{\partial\Omega}e((q_1,T^*_{B_1}(\mathcal B^*\bar\Phi'V_0))-(q_4,P^*_{-A^*_2}(\mathcal A\Phi'V_0))e^{\tau(\Phi-\bar\Phi)}d\sigma=c_3+o(\frac 1\tau)\quad \mbox{as}\,\,\tau\rightarrow +\infty.
\end{equation}
and
\begin{equation}\label{gop4}
\int_{\partial\Omega}e((q_2,P_{A_1}(\mathcal B\Phi'\tilde V_0))-(q_3,T^*_{-B^*_2}(\mathcal A^*\bar \Phi'\tilde U_0))e^{\tau(\Phi-\bar\Phi)}d\sigma=c_4+o(\frac 1\tau)\quad \mbox{as}\,\,\tau\rightarrow +\infty.
\end{equation}
Then using (\ref{gop1})-(\ref{gop4}) and using Theorem 7.7.1 of \cite{Her},
we have

\begin{equation}\label{vovik1}
((\nu_1-i\nu_2)\mathcal AU_0,V_0)(x^*)-2(q_1,T^*_{B_1}(\mathcal B^*\bar\Phi'V_0))(x^*)-2(q_4,P^*_{-A^*_2}(\mathcal A\Phi'V_0))(x^*)=0
\end{equation}
and

\begin{equation}\label{vovik2}
((\nu_1+i\nu_2)\mathcal BU_0,V_0)(x^*)-2(q_2,P_{A_1}(\mathcal B\Phi'\tilde V_0))(x^*)-2(q_3,T^*_{-B^*_2}(\mathcal A^*\bar \Phi'\tilde U_0))(x^*)=0.
\end{equation}

By Proposition  \ref{nikita} for any an arbitrary vectors $\vec z_1, \vec z_2, \vec z_3, \vec z_4$ we can choose function $q_j$ in such a way, that $q_j(x^*)=\vec z_j.$
Therefore (\ref{vovik1}) and (\ref{vovik2}) imply (\ref{ppp}).
In order to prove (\ref{pppp}) we  consider the complex geometric optics solution constructed with the phase function $\tilde \Phi$ instead of $\Phi$ and  the pair $(\frac{1}{\tilde \Phi'}U_0, \frac{1}{\overline{\tilde \Phi'}}\tilde U_0)$ instead of $(U_0,\tilde U_0).$ By Proposition \ref{lodka}  the equality (\ref{zaika}) holds true. Since there are no critical points of the function $\tilde\Phi$ on $\bar\Omega$  the terms $I_\pm , K_\pm, J_\pm$ are equal to zero. Then applying the   theorem 7.7.5 of \cite{Her} we have

\begin{equation}\label{vovik1}
((\nu_1-i\nu_2)\mathcal AU_0,V_0)(x^*)-2(q_1,T^*_{B_1}(\mathcal B^*V_0))(x^*)-2(q_4,P^*_{-A^*_2}(\mathcal AV_0))(x^*)=0
\end{equation}
and

\begin{equation}\label{vovik2}
((\nu_1+i\nu_2)\mathcal BU_0,V_0)(x^*)-2(q_2,P_{A_1}(\mathcal B\tilde V_0))(x^*)-2(q_3,T^*_{-B^*_2}(\mathcal A^*\tilde U_0))(x^*)=0.
\end{equation}
By Proposition  \ref{nikita} for any an arbitrary vectors $\vec z_1, \vec z_2, \vec z_3, \vec z_4$ we can choose function $q_j$ in such a way, that $q_j(x^*)=\vec z_j.$
Therefore (\ref{vovik1}) and (\ref{vovik2}) imply (\ref{pppp}).
 $\blacksquare$

Using Proposition \ref{knopka} we prove the following:

\begin{proposition}\label{loop}
Under assumptions  of Proposition \ref{lodka} the following equality holds true
\begin{equation}\label{donka}
P_{-A_2^*}^*(\mathcal AU_0\Phi')= \Phi'P_{-A_2^*}^*(\mathcal A U_0),\quad T_{B_1}(\mathcal B^*\bar\Phi'V_0)=\overline{\Phi}' T_{B_1}(\mathcal B^* V_0)
\end{equation}
\begin{equation}
P_{A_1^*}^*(\mathcal A^*\tilde V_0\Phi')=\Phi'P_{A_1^*}^*(\mathcal A^*\tilde V_0),\quad T^*_{-B^*_2}(\mathcal B\bar\Phi'\tilde U_0)=\overline{\Phi}' T^*_{-B_2^*}(\mathcal B^*\tilde U_0).
\end{equation}
\end{proposition}

{\bf Proof.} We prove the first equality in (\ref{donka}). The proof of the remaining three equalities is the same. By (\ref{ppp}) and (\ref{pppp})
$$
P_{-A_2^*}^*(\mathcal AU_0\Phi')=\Phi'P_{-A_2^*}^*(\mathcal AU_0)=0 \quad \mbox{on}\,\,\tilde \Gamma .
$$
We set $r_1=P_{-A_2^*}^*(\mathcal AU_0\Phi')$ and $r_2=\Phi'P_{-A_2^*}^*(\mathcal AU_0).$
Then functions $r_j$ satisfy
$$
-2\partial_{\bar z}r_j-A_2^*r_j=\mathcal AU_0\Phi'\quad \mbox{in}\,\,\Omega, \quad r_j\vert_{\tilde \Gamma}=0\quad j\in\{1,2\}.
$$
By the uniqueness of the Cauchy problem for the $\partial_{\bar z}$- operator
we have that $r_1=r_2.$  The proof of Proposition \ref{loop} is complete.
$\blacksquare$

We use Proposition \ref{loop} to simplify the formula (\ref{zaika}).

\begin{proposition}\label{zanuda1}
Under conditions of Proposition \ref{lodka}  we have
\begin{eqnarray}\label{victorykk}
-(2\partial_z\mathcal AU_0, V_0)(\tilde x)-(2\mathcal AU_0, \partial_zV_0)(\tilde x) +(2\mathcal B\partial_{\bar z} U_0, V_0)(\tilde x)\\-((Q_1(1)U_0, T^*_{B_1}(\mathcal B^*V_0))+(Q(2)V_0, P^*_{-A^*_2}(\mathcal AU_0)))(\tilde x)+(\mathcal Q U_0,V_0)(\tilde x)+I_{+,\Phi}(\tilde x)=0\quad\mbox{in}\,\,\,\Omega,\nonumber
\end{eqnarray}
where function $\Phi$ satisfies (\ref{mika}) and
\begin{equation}\label{gromila}
\mbox{Im}\,\Phi(\widetilde x)\notin \{\mbox{Im}\,
\Phi(x); \thinspace x\in \mathcal H\setminus
\{\widetilde{x}\}\}.
\end{equation}
\end{proposition}

{\bf Proof.} By Proposition \ref{lodka} equality (\ref{zaika}) holds true. Thanks to (\ref{gromila}) and Proposition \ref{knopka} we can write it as
\begin{equation}\label{gromiko}
(J_++I_++K_+)(\tilde x)-\pi(\mathcal L (q_1,T^*_{B_1}(\mathcal B^*\bar\Phi'V_0))+\mathcal L(q_4, P_{-A_2}^*(\mathcal A\Phi'U_0))+I_{+,\Phi}(\tilde x)=0,
\end{equation}
where $\mathcal Lu=-\frac{\partial_{zz}^2u(\widetilde x)}{2\Phi''(\tilde x)}+\frac{\partial^2_{\overline z\overline z}u(\widetilde x)}{2\bar\Phi''(\tilde x)}.$
Let us transform transform  the function $K_+$ which is given by (\ref{zubilo}).By Propositions \ref{osel} and \ref{loop}, we obtain

\begin{eqnarray}\label{lenta12}
\frak F_\tau(q_1 , T^*_{B_1}B_1^*\mathcal A^* V_0-\mathcal A^*V_0+2T_{B_1}^*\partial_z \mathcal B^* V_0+T_{B_1}^*(\mathcal B^*(A^*_2V_0-2\tau\bar\Phi'V_0)))=\nonumber\\
-\frak F_\tau(q_1 ,T_{B_1}^*(\mathcal B^*2\tau\bar\Phi'V_0))=-\frak F_\tau(q_1 ,2\tau\bar\Phi'T_{B_1}^*(\mathcal B^*V_0))=\nonumber\\
-\frac{ \pi}{ 2\tau}(2\partial_{\bar z} q_1,T_{B_1}^*(\mathcal B^*V_0))(\tilde x)=-\frac {\pi}{ 2\tau}(Q_1(1)U_0,T_{B_1}^*(\mathcal B^*V_0))(\tilde x)
\end{eqnarray}

and

\begin{eqnarray}\label{lenta22}
-2\frak F_{\tau}(P_{-A_2^*}^*(\mathcal A (\partial_zU_0+\tau\Phi' U_0))+\mathcal B \partial_{\bar z} U_{0,\tau}, q_4)=-2\frak F_{\tau}(P_{-A_2^*}^*(\mathcal A\tau\Phi' U_0)), q_4)=\\
-2\frak F_{\tau}(\tau\Phi'(P_{-A_2^*}^*(\mathcal A U_0)), q_4)=\frac{\pi}{ 2\tau} (P_{-A_2^*}^*(\mathcal A U_0), 2\partial_z q_4)(\tilde x)=\frac{\pi}{ 2\tau} (P_{-A_2^*}^*(\mathcal A U_0), Q_2(2)V_0)(\tilde x).\nonumber
\end{eqnarray}
By (\ref{lenta12}) and (\ref{lenta22})
\begin{equation}\label{lenta3}
K_+=-\frac{ \pi}{ 2\tau}(Q_1(1)U_0,T_{B_1}^*(\mathcal B^*V_0))(\tilde x)+\frac{\pi}{ 2\tau} (P_{-A_2^*}^*(\mathcal A U_0), Q_2(2)V_0)(\tilde x).
\end{equation}

Substituting into equality (\ref{gromiko}) the right hand side of  formula (\ref{lenta3}) we obtain (\ref{victorykk}).
The proof of the proposition is complete.
$\blacksquare$

\bigskip
\section{\bf Step 5: End of the proof.}
\bigskip
Suppose that for the operators  $L_{\mu_1,\gamma_1}(x,D)$  and $L_{\mu_2,\gamma_2}(x,D)$  given by (\ref{zoika})  and the Dirichlet-to-Neumann maps  $\Lambda_{\mu_j,\gamma_j}$ given by (\ref{2}) are the same.
 Then,  by Proposition \ref{leika}   the Dirichlet-to-Neumann maps for the operators $L_1(x,D)$ and $L_2(x,D)$ given by  formula (\ref{kinovolt})  are the same.
 Let $u_1$ be the complex geometric optics solution given by (\ref{zad}) constructed for the operator $L_1(x,D).$
There exists a  function   $ u_2$ be a solution to the following boundary value
problem:
$$
{ L}_{2}(x,D) u_2=0\quad \mbox{in}\,\,\Omega,\quad
 \mathcal B(x,D)(u_1-u_2)\vert_{\partial\Omega}=0, \quad
\mathcal  R(x,D) (u_1-u_2)=0\quad \mbox{on}\,\,\tilde \Gamma.
$$
Let $\eta$ be a function such that
\begin{equation}\label{napoleon}
\eta\in C^\infty_0(\Omega),\quad  \eta\vert_{\mathcal O}=1,
\end{equation} where $\mathcal O$ is some open set such that  $\mathcal H\subset O\subset \Omega.$
The operator $L_{1}(x,s,D)=e^{s\eta}L_1(x,D) e^{-s\eta}$ has the same Dirichlet-to-Neumann map as the operator $L_1(x,D).$
Then the function $\tilde u_1=e^{s\eta}u_1$ solves the boundary value problem
\begin{equation}\label{(2.1I)}
{ L}_{1}(x,s,D)\tilde u_1=0\quad \mbox{in}\,\,\Omega,\quad
\mathcal B(x,D)(\tilde u_1-u_2)\vert_{\partial\Omega}=0, \quad
\mathcal R(x,D) (\tilde u_1-u_2)=0\quad \mbox{on}\,\,\tilde \Gamma.
\end{equation}

Setting $u=\tilde u_1-u_2$, $\mathcal A_s=A_1-2s\partial_{\bar z}\eta-A_2, \mathcal A=\mathcal A_0$ $\mathcal B_s=B_1-2s\partial_z\eta-B_2, \mathcal B=\mathcal B_0$ and $\mathcal Q_s=Q_1-Q_2-s^2\vert\nabla\eta\vert^2+s\Delta\eta+2sA_2\partial_z\eta+2sB_2\partial_{\bar z}\eta, \mathcal Q=\mathcal Q_0$ we have
\begin{equation}
{L}_2(x,{D})u +2\mathcal A_s\partial_z u_1
+2\mathcal B_s\partial_{\overline z} u_1
+\mathcal Q_s\tilde u_1=0 \quad \mbox{in}~ \Omega    \label{mn}
\end{equation}
and
\begin{equation}\label{mn1}
\mathcal B(x,D)u \vert_{\partial\Omega} =0, \quad \mathcal R(x,D)u
\vert_{\widetilde \Gamma} =0.
\end{equation}
Let $v$ be a function given by  (\ref{-3}).  Taking the scalar
product of (\ref{mn}) with $ v$  in $L^2(\Omega)$ and using
(\ref{-4}) and (\ref{mn1}), we obtain
\begin{equation}\label{ippolit}
0=\frak G(\tilde u_1,v)= \int_{\Omega}(2\mathcal A_s\partial_z\tilde
u_1 +2\mathcal B_s\partial_{\overline z}\tilde  u_1 +\mathcal Q_s\tilde u_1,
v) dx.
\end{equation}

\begin{proposition}\label{kauk}
Let $\widetilde u_1=e^{s\eta}u_1,$ where $u_1$  is given by  (\ref{zad})
and $v$ is given by  (\ref{-3}). Then the following asymptotics holds true
$$
\frak G(\tilde u_1,v)=\int_{\Omega}(2\mathcal A_s\partial_z
(e^{s\eta} U) +2\mathcal B_s\partial_{\overline z}(e^{s\eta} U) +\mathcal Q_s (e^{s\eta} U),
V^*) dx+o(\frac 1\tau)\quad \mbox{as}\quad\tau\rightarrow +\infty.
$$
\end{proposition}
{\bf Proof.}
Since the form $\frak G(\cdot,\cdot) $ is bilinear  in order to prove the statement of this proposition it suffices to show that
\begin{equation}\label{-15}
\frak G(e^{\tau\varphi}\tilde u_{-1},  V^*)=\frak G(e^{\tau\varphi}\tilde u_{-1}, e^{-\tau\varphi} v_{-1})=\frak G( e^{s\eta} U,  e^{-\tau\varphi}v_{-1})=o(\frac 1\tau)\quad \mbox{as}\,\,\tau\rightarrow +\infty .
\end{equation}

Obviously, by (\ref{mimino11}) and (\ref{Amimino11}), we see that
\begin{equation}
\frak G(e^{\tau\varphi}\tilde u_{-1},e^{-\tau\varphi}v_{-1})=o(\frac 1\tau
)\quad\mbox{as}\,\,\tau\rightarrow +\infty.
\end{equation}

Let $\chi\in C_0^\infty(\Omega)$ satisfy
$\chi\vert_{\Omega\setminus \mathcal O_{\frac \epsilon 2}}=1.$
By (\ref{mimino11}), we have
\begin{eqnarray}\label{llg1}
\frak G(e^{\tau\varphi}\tilde u_{-1},  V^*)=\frak
G(e^{\tau\varphi}\tilde u_{-1}, \chi V^*)+o(\frac 1\tau) \nonumber\\
= \int_\Omega (2\mathcal A_s\partial_z(e^{\tau\varphi}\tilde u_{-1})+2\mathcal B_s
\partial_{\overline z}(e^{\tau\varphi}\tilde u_{-1}),
\chi V^*)dx + o(\frac 1\tau) \nonumber\\
= \int_\Omega
(2\mathcal A_s\partial_z(e^{\tau\varphi}\tilde u_{-1}),
{\chi \tilde V_0e^{-\tau\overline\Phi}})+(2\mathcal B_s\partial_{\overline
z}(e^{\tau\varphi}\tilde u_{-1}),{\chi
V_0e^{-\tau\Phi}})dx+o(\frac 1\tau).
\end{eqnarray}
Let functions $w_4,w_5$ solve the equations $(-\partial_{\overline
z}+(B^*_1-2s\partial_{ z}\eta))w_4=2{\mathcal A}_s^*V_0$ and $(-\partial_{
z}+(A^*_1-2s\partial_{\bar z}\eta))w_5=2{\mathcal B}_s^*\tilde V_0.$

Taking the scalar product of equation (\ref{zad2!}) and the function
$\chi(w_5e^{\tau\Phi}+w_4e^{\tau\overline\Phi})$, after integration by parts we
obtain
\begin{equation}\label{llg}
\int_\Omega((2\partial_z
(e^{\tau\varphi}\tilde u_{-1})+(A_1+2s\partial_{\bar z}\eta)(e^{\tau\varphi}\tilde u_{-1}),
2\mathcal A_s^*{\widetilde
V_0e^{-\tau\overline\Phi}})
\end{equation}
$$
+(2\partial_{\overline z}
(e^{\tau\varphi}\tilde u_{-1})+(B_1-2s\partial_z\eta)(e^{\tau\varphi}\tilde u_{-1}),2{\mathcal B}^*_s
{\tilde V_0e^{-\tau\Phi}}))dx=o(\frac{1}{\tau}).
$$
By (\ref{llg1}) and (\ref{llg}), we obtain the first equality in
(\ref{-15}). The proof of the second equality in (\ref{-15}) is the
same and involves the estimate (\ref{Amimino11}).
$\blacksquare$

Thanks to Proposition \ref{kauk}, the statements of Propositions \ref{lodka}
- \ref{loop} hold true.

\begin{proposition}\label{zanuda} Let sequence of function $\Phi_\epsilon$ given by Proposition \ref{Proposition -1}.
For the functions $I_{\pm,\Phi_\epsilon}$ given by (\ref{lin1}) and (\ref{lin2}) we have
\begin{equation}\label{noga}
I_{\pm,\Phi_\epsilon}(\tilde x_\epsilon)\equiv 0.
\end{equation}
\end{proposition}
{\bf Proof.} We prove this statement for the function $I_{+,\Phi_\epsilon}.$
The proof for the function $I_{-,\Phi_\epsilon}$ is the same. By Proposition \ref{zanuda1}  equality (\ref{victorykk}) holds true. Next we observe that
\begin{equation}\label{volga1}
T^*_{B_1}(\mathcal B^* V_0)=V_0+p_1, \quad  p_1\in \mbox{Ker}\, T^*_{B_1}
\end{equation}
and
\begin{equation}\label{volga2}
P^*_{-A^*_2}(\mathcal A U_0)=U_0+p_2, \quad  p_2\in \mbox{Ker}\, P^*_{-A^*_2}.
\end{equation}
By Proposition \ref{knopka}
\begin{equation}\label{vilka}
V_0+p_1=U_0+p_2=0\quad \mbox{on}\,\,\tilde \Gamma.
\end{equation}
Let  function $\eta$ satisfies (\ref{napoleon}).
The operator $L_{1,s}(x,D)=e^{s\eta}L_1(x,D) e^{-s\eta}$ has the same Diriclet-to-Neumann map as the operator $L_1(x,D).$ For the operator $L_{1,s}(x,D)$ the equation (\ref{victory1}) has the form
\begin{eqnarray}\label{victory1}
-(2\partial_z\mathcal A_sU_0, V_0)e^{s}-(2\mathcal A_sU_0, \partial_zV_0)e^{s} +(2\mathcal B_s\partial_{\bar z} U_0, V_0)e^{s}\\
-((Q_1(1)U_0, T^*_{B_1}(\mathcal B^*_se^{s\eta}V_0))+(Q(2)V_0, P^*_{-A^*_2}(\mathcal A_se^{s\eta}U_0)))+e^{s}(\mathcal Q_s U_0,V_0)+I_{+,\Phi_\epsilon}=0.\nonumber
\end{eqnarray}
We claim that
\begin{equation}\label{v}
T^*_{B_1}(\mathcal B^*_s V_0)=e^{s\eta}(V_0+p_1), \quad  p_1\in \mbox{Ker}\, T^*_{B_1}
\end{equation}
and
\begin{equation}\label{v1}
P^*_{-A^*_2}(\mathcal A_s U_0)=e^{s\eta}(U_0+p_2), \quad  p_2\in \mbox{Ker}\, P^*_{-A^*_2}.
\end{equation} where functions $p_1$ and $p_2$ are given by (\ref{volga1}) and (\ref{volga2}) respectively. The direct computations imply that
\begin{equation}\label{robot2}
(-\partial_z+B_1^*)(e^{s\eta}(V_0+p_1))=e^{s\eta}(\mathcal B^*_s V_0)\quad \mbox{and} \quad (-\partial_{\bar z}+A_2)(e^{s\eta}(U_0+p_2))=e^{s\eta}(\mathcal A_s U_0).
\end{equation}
By Proposition  \ref{knopka}
\begin{equation}\label{robot1} T^*_{B_1}(\mathcal B^*_s V_0)=P^*_{-A^*_2}(\mathcal A_s U_0)=0\quad\mbox{on}\quad\tilde \Gamma.
\end{equation}
On the other hand by (\ref{vilka}) we have
$$
e^{s\eta}(V_0+p_1)=e^{s\eta}(U_0+p_2)=0\quad \mbox{on}\,\,\tilde \Gamma.
$$
The by uniqueness of solution  for the Cauchy problem for $\partial_z$ operator we have (\ref{v}) and (\ref{v1}).
Using  (\ref{robot1}) and (\ref{robot2}) we rewrite equation (\ref{victorykk}) as

\begin{eqnarray}\label{victory1}
-(2\partial_z\mathcal AU_0, V_0)e^{s}-(2\mathcal AU_0, \partial_zV_0)e^{s} +(2\mathcal B\partial_{\bar z} U_0, V_0)e^{s}\\
-e^{s}((Q_1(1)U_0, T^*_{B_1}(\mathcal B^*V_0))+e^{s}(Q(2)V_0, P^*_{-A^*_2}(\mathcal AU_0)))+e^{s}(\mathcal Q U_0,V_0)+I_{+,\Phi_\epsilon}(\tilde x)=0.\nonumber
\end{eqnarray}
This imply (\ref{noga}).
$\blacksquare$

Using (\ref{noga}) and the fact that by (\ref{bobik2})  for any $x$ from $\Omega$  exists a sequence of $x_\epsilon$  converging to $x$ we rewrite  the equation (\ref{victory}) as
\begin{eqnarray}\label{victory5}
-(2\partial_z\mathcal AU_0, V_0)-(2\mathcal AU_0, \partial_zV_0) +(2\mathcal B\partial_{\bar z} U_0, V_0)\nonumber\\-((Q_1(1)U_0, T^*_{B_1}(\mathcal B^*V_0))+(Q(2)V_0, P^*_{-A^*_2}(\mathcal AU_0)))+(\mathcal Q U_0,V_0)=0\quad\mbox{in}\,\,\,\Omega.
\end{eqnarray}
By Proposition \ref{nikita}  for each point $\tilde x$ from $\Omega$ one can construct such a function $U_0,V_0$ satisfying (\ref{-55!!}),(\ref{xoxo1u}), (\ref{ll1}), (\ref{iiii}) such that
$$
U^{(k)}_0(\tilde x)=\vec e_k,\quad V^{(\ell)}_0(\tilde x)=\vec e_\ell\quad k,\ell\in \{1,2\}.
$$
Then for each $\tilde x$ there exists positive $\delta(\tilde x)$ such that the matrices $\{U^{(j)}_{0,i}\}$ and $\{V^{(j)}_{0,i}\}$ are invertible for any $x\in \overline{B(\tilde x,\delta(x))}. $ From the covering of $\Omega$ by such a balls we take the finite subcovering $\bar \Omega\subset \cup_{k=1}^{\tilde N} B(x_k,\delta_k).$
Then from (\ref{victory5}) we have the
differential inequality
\begin{equation}\label{london}
\vert \partial _z\mathcal A_{ij}\vert \le \mathcal C(x)(\sum_{k=1}^N\vert T^*_{B_1}(\mathcal B^*V_0(k))\vert+ \vert P^*_{-A^*_2}(\mathcal AU_0(k))\vert+\vert\mathcal A\vert+\vert\mathcal B\vert +\vert\mathcal Q\vert)\quad\mbox{in}\,\,\Omega, \quad \forall i,j\in \{1,2\}.
\end{equation}
We set $\rho=(\rho_1,\rho_2,\rho_3)$ where the function $\rho_i$ are defined by (\ref{gnom}). Let point $\hat x$ belongs to $\mbox{supp}\rho.$
Using (\ref{megaphon}) we rewrite (\ref{london}) as
\begin{equation}\label{oblom}
\vert \Delta \rho\vert \le \mathcal C(x)(\sum_{k=1}^N\vert T^*_{B_1}(\mathcal B^*V_0(k))\vert+ \vert P^*_{-A^*_2}(\mathcal AU_0(k))\vert +\vert \nabla \rho\vert+\vert\rho\vert )\quad\mbox{in}\,\,\Omega.
\end{equation}
Let $\tilde\gamma$ be a curve, without self-intersections  which pass through the point $\hat x$ and couple points $x_1, x_2$ from $\tilde\Gamma$  in such a way that the set  $\tilde \gamma\cap\partial\Omega\setminus\{x_1,x_2\}$ is empty. Denote by $\Omega_1$ a domain bounded by $\tilde\gamma$ and  and part of
$\partial\Omega$ located between points $x_1$ and $x_2.$ Then we set
$\Omega_{1,\epsilon}=\{x; \thinspace dist(\Omega_1,x)<\epsilon\}.$
Let $\phi_0$ be a function such that
\begin{equation}\label{indigo}
\nabla \phi_0(x)\ne 0\quad\mbox{in}\,\,\Omega_1,\quad \partial_{\tilde\nu}\phi_0\vert_{\tilde \gamma}\le \alpha' <0,\quad \phi_0\vert_{\tilde \gamma}=0.
\end{equation}
where $\tilde \nu$ is the outward normal derivative to $\Omega_1$
and $\mu_\epsilon$ be a function such that
$$
 \mu_\epsilon\in C_0^\infty(\Omega_{1,\epsilon}), \quad \mu_\epsilon=1\quad \quad\mbox{in}\,\,\Omega_1.
 $$ We set $\rho_\epsilon=\mu_\epsilon \rho.$
From (\ref{oblom}) we have
$$
\vert \Delta \rho_{\epsilon}\vert \le \mathcal C(x)(\sum_{k=1}^N\vert \mu_\epsilon  T^*_{B_1}(\mathcal B^*V_0(k))\vert+ \vert\mu_\epsilon P^*_{-A^*_2}(\mathcal AU_0(k))\vert +\mu_\epsilon(\vert \nabla \rho\vert+\vert\rho\vert)+\vert [\mu_\epsilon,\Delta]\rho\vert )\quad\mbox{in}\,\,\Omega_{1,\epsilon},\quad \rho_\epsilon\vert_{\partial\Omega_{1,\epsilon}}=0.
$$

Set $\psi_0=e^{\lambda\phi_0}$ with positive $\lambda$ sufficiently large.
Applying the Carleman estimate to the above inequality we have
\begin{eqnarray}\label{volk1}
\int_{\Omega_{1,\epsilon}} e^{2\tau\psi_0}(\tau\vert \nabla \rho_\epsilon\vert^2+\tau^3\vert \rho_\epsilon\vert^2) dx\le C \int_{\Omega_{1,\epsilon}} (\sum_{k=1}^N\vert \mu_\epsilon  T^*_{B_1}(\mathcal B^*V_0(k))\vert^2\nonumber\\+ \vert\mu_\epsilon P^*_{-A^*_2}(\mathcal AU_0(k))\vert^2 +\mu_\epsilon^2
(\vert \nabla \rho\vert+\vert\rho\vert^2)+\vert [\mu_\epsilon,\Delta]\rho\vert^2 ) e^{2\tau\psi_0}dx\quad\forall\tau\ge \tau_0.
\end{eqnarray}

By the Carleman estimate for the operator $\partial_z$  there exist $C$ and $\tau_0$ independent of $\tau $ such that
\begin{equation}\label{volk2}\int_{\Omega_{1,\epsilon}}\vert \mu_\epsilon  T^*_{B_1}(\mathcal B^*V_0(k))\vert^2 e^{2\tau\psi_0}dx\le C\int_{\Omega_{1,\epsilon}}(\vert[\mu_\epsilon ,\partial_z] T^*_{B_1}(\mathcal B^*V_0(k))\vert^2+\vert\mathcal B^*V_0(k)\vert^2) e^{2\tau\psi_0}dx
\end{equation}
and
\begin{equation}\label{volk3}\int_{\Omega_{1,\epsilon}}\vert \mu_\epsilon   P^*_{-A^*_2}(\mathcal AU_0(k))\vert^2 e^{2\tau\psi_0}dx\le C\int_{\Omega_{1,\epsilon}}(\vert[\mu_\epsilon ,\partial_{\bar z}]  P^*_{-A^*_2}(\mathcal AU_0(k))\vert^2+\vert (\mathcal AU_0(k))\vert^2) e^{2\tau\psi_0}dx \quad \forall\tau\ge\tau_0
\end{equation} for all $\tau\ge\tau_0.$
Combining (\ref{volk1}),  (\ref{volk2}) and (\ref{volk3}) we obtain
\begin{eqnarray}\label{volk4}
\int_{\Omega_{1,\epsilon}} e^{2\tau\psi_0}(\tau\vert \nabla \rho_\epsilon\vert^2+\tau^3\vert \rho_\epsilon\vert^2) dx\le C \int_{\Omega_{1,\epsilon}}(\sum_{k=1}^N\vert[\mu_\epsilon ,\partial_{\bar z}]  P^*_{-A^*_2}(\mathcal AU_0(k))\vert^2\\+\vert\mu_\epsilon (\mathcal AU_0(k))\vert^2+\vert[\mu_\epsilon ,\partial_z] T^*_{B_1}(\mathcal B^*V_0(k))\vert^2+\vert\mu_\epsilon\mathcal B^*V_0(k)\vert^2+
\vert [\mu_\epsilon,\Delta]\rho\vert^2 ) e^{2\tau\psi_0}dx\nonumber\\
 \le C \int_{\Omega_{1,\epsilon}}(\sum_{k=1}^N\vert[\mu_\epsilon ,\partial_{\bar z}]  P^*_{-A^*_2}(\mathcal AU_0(k))\vert^2\nonumber
 + \vert \nabla \rho_\epsilon\vert^2+\vert \rho_\epsilon\vert^2)\\+\vert[\mu_\epsilon ,\partial_z] T^*_{B_1}(\mathcal B^*V_0(k))\vert^2+
 \vert [\mu_\epsilon,\Delta]\rho\vert^2 ) e^{2\tau\psi_0}dx \quad \forall\tau\ge\tau_0.\nonumber
\end{eqnarray} For all sufficiently large $\tau$  the term $\int_{\Omega_{1,\epsilon}}(\vert \nabla \rho_\epsilon\vert^2+\vert \rho_\epsilon\vert^2)e^{2\tau\psi_0}dx$ absorbed by the integral on the left hand side. Moreover, thanks to the choice of the  function $\mu_\epsilon,$ we have supports of coefficients for the operators $[\mu_\epsilon ,\partial_z], [\mu_\epsilon ,\partial_{\bar z}]$ and $[\mu_\epsilon,\Delta]$ are located in the domain $\Omega_{1,\epsilon}\setminus\Omega_{1,\frac{\epsilon}{2}}.$

\begin{eqnarray}\label{volk4}
\int_{\Omega_{1,\epsilon}} e^{2\tau\psi_0}(\tau\vert \nabla \rho_\epsilon\vert^2+\tau^3\vert \rho_\epsilon\vert^2) dx\le C\int_{\Omega_{1,\epsilon}\setminus (\Omega_{1,\frac{\epsilon}{2}}\cup\Omega_1)}
(\sum_{k=1}^N\vert[\mu_\epsilon ,\partial_{\bar z}]  P^*_{-A^*_2}(\mathcal AU_0(k))\vert^2\nonumber\\
 +\vert[\mu_\epsilon ,\partial_z] T^*_{B_1}(\mathcal B^*V_0(k))\vert^2+
\vert [\mu_\epsilon,\Delta]\rho\vert^2 ) e^{2\tau\psi_0}dx\quad \forall\tau\ge\tau_1.
 \end{eqnarray}
 By (\ref{indigo}) for all sufficiently small positive $\epsilon$ there exists a positive constant $\alpha<1$ such that
 \begin{equation}\label{soroka}
 \tilde \psi_0(x) <\alpha\quad\mbox{on} \quad\Omega_{1,\epsilon}\setminus(\Omega_{1,\frac{\epsilon}{2}}\cup \Omega_1).
 \end{equation}
  Since $\hat x\in \mbox{supp}\, \rho\cap \tilde\gamma$  and  thanks to the fact $\partial_{\tilde\nu}\phi_0\vert_{\tilde \gamma}\le \alpha ' <0$ there exists $\kappa>0$ such that
 \begin{equation}\label{volk6}\kappa e^\tau\le \int_{\Omega_{1,\epsilon}} e^{2\tau\psi_0}(\tau\vert \nabla \rho_\epsilon\vert^2+\tau^3\vert \rho_\epsilon\vert^2) dx \quad \forall\tau\ge\tau_1.
 \end{equation}
  By (\ref{soroka}) we can estimate the right hand side of the inequality (\ref{volk4}) as
  \begin{eqnarray}\label{volk5}
\int_{\Omega_{1,\epsilon}\setminus (\Omega_{1,\frac{\epsilon}{2}}\cup\Omega_1)}
(\sum_{k=1}^N\vert[\mu_\epsilon ,\partial_{\bar z}]  P^*_{-A^*_2}(\mathcal AU_0(k))\vert^2\nonumber\\
 +\vert[\mu_\epsilon ,\partial_z] T^*_{B_1}(\mathcal B^*V_0(k))\vert^2+
\vert [\mu_\epsilon,\Delta]\rho\vert^2 ) e^{2\tau\psi_0}dx\le C e^{\alpha\tau}\quad \forall\tau\ge\tau_1.
\end{eqnarray}
Using (\ref{volk6}) and (\ref{volk5}) in (\ref{volk4}) we obtain
$$
\kappa e^\tau \le C e^{\alpha\tau}\quad \forall\tau\ge\tau_1.
$$
Since $\alpha<1$ we arrived to the contradiction.
 $\blacksquare$


\begin{thebibliography}{99} %

\bibitem{POM} P. Caro, P. Ola and S. Mikko, {\it Inverse boundary value problem for Maxwell equations  with local data} Com. P.D.E., {\bf 34}, (2009), 1425-1464.



\bibitem{C}  A. P.\ Calder\'on, \textit{On an inverse boundary value
problem,} in \emph{Seminar on Numerical Analysis and its
Applications to Continuum Physics}, 65--73, Soc. Brasil. Mat., R\'io
de Janeiro, 1980.


\bibitem{DL} R. Dautray and J.-L. Lions,
\textit{Mathematical Analysis and Numerical Methods for Science and
Technology vol. 3}, Springer-Verlag, Berlin, 2000.

\bibitem{Her} L. H\"ormander, \textit{ The analysis of linear partial differential operators I,} Springer-Verlag, Berlin, 1980.

\bibitem{IUY} O. Imanuvilov, G. Uhlmann and M. Yamamoto,
\textit{The Calder\'on problem with partial data in two dimensions},
J. Amer. Math. Soc., {\bf 23} (2010), 655-691.

\bibitem{IUY3} O. Imanuvilov, G. Uhlmann and M. Yamamoto,
\textit{Inverse boundary value problem by partial data for the Neumann-to-Dirichlet map in two dimensions,}
arXiv:1210.1255v1

\bibitem{IY} O. Imanuvilov and M. Yamamoto,
\textit{Inverse problem by Cauchy data on an arbitrary sub-boundary for systems of elliptic equations},
Inverse Problems , {\bf 28} (2013), 095015.

\bibitem{IYAM} O. Imanuvilov and M. Yamamoto,
\textit{Uniqueness for inverse boundary value problems by Dirichlet-to-Neumann
map on subboundarires, } Milan J. Math., {\bf 81} (2013), 187-258.

\bibitem{IY2} O. Imanuvilov and M. Yamamoto,
\textit{Calder\'on problem for Maxwell's equations in two dimensions},
arXiv:1403.7596






\bibitem{J} J. D. Jackson,
\textit{Classical Electrodynamics}, John Wiley \& Sons, New York, 1962.

\bibitem{OPS} P. Ola, L. P\" aiv\"arinta and E. Somersalo,
{\it An inverse boundary value problem in electrodynamics,}
Duke . Math. J.,
\textbf {70} (1993), 617-653

\bibitem{SU} J.~Sylvester and G.~Uhlmann,
\textit{A global uniqueness theorem for an inverse boundary value
problem}, Ann. of Math., \textbf{125}  (1987), 153--169.

\bibitem{VE}  I.\ Vekua, \textit{Generalized Analytic Functions},
Pergamon Press, Oxford, 1962.

\bibitem{Wendland} W.\ Wendland, \textit{Elliptic Systems in the Plane},
Pittman, London, 1979.

\end{thebibliography}
\end{document}